\documentstyle[eqsecnum,epsf,aps,prb]{revtex}
\title{Electron-electron interactions, quantum 
Coulomb gap, and dynamical scaling near integer quantum Hall transitions }
\author{Ziqiang Wang and Shanhui Xiong}
\address{Department of Physics, Boston College, Chestnut Hill, MA 
02467}

\date{\today}
\begin{document}
\maketitle
\renewcommand{\thesubsubsection}{\arabic{subsubsection}}
\newcommand{\pprl}{Phys. Rev. Lett. \ } 
\newcommand{\pprb}{Phys. Rev. {B}} 
\newcommand{\be}{\begin{equation}}
\newcommand{\bea}{\begin{eqnarray}}
\newcommand{\eea}{\end{eqnarray}}
\newcommand{\ee}{\end{equation}}
\newcommand{\br}{{\bf r}}
\newcommand{\brp}{{\bf r^\prime}}
\newcommand{\ov}{\overline}
\newcommand{\ba}{\begin{array}}
\newcommand{\ea}{\end{array}}
\newcommand{\del}{\delta}
\newcommand{\al}{\alpha}
\newcommand{\non}{\nonumber}
\newcommand{\dl}{\delta}
\newcommand{\k}{\kappa}
\newcommand{\e}{\epsilon}
\newcommand{\zt}{z_{\rm T}}
\newcommand{\Tr}{{\rm Tr}}
\def\sxx{\sigma_{xx}}
\def\sxxc{\sigma_c}
\def\sxy{\sigma_{xy}}
\def\de{\Delta E}
\def\ef{E_F}
\def\dxx{D}
\def\dxy{D_{xy}}
\newcommand{\nuzt}{\tilde{\nu}_0}
\newcommand{\dt}{\partial_t}
\newcommand{\jotau}{j_0^\tau}
\newcommand{\jmu}{j_\mu}
\newcommand{\bpsi}{\bar\psi}
\newcommand{\dtau}{\partial_\tau}
\newcommand{\da}{\partial_\alpha}
\newcommand{\db}{\partial_\beta}
\newcommand{\dmu}{\partial_\mu}
\newcommand{\dnu}{\partial_\nu}
\newcommand{\vimp}{V}
\newcommand{\ve}{\varepsilon}
\newcommand{\ave}{\vert\varepsilon\vert}
\newcommand{\vov}{\vert\omega\vert}
\newcommand{\vnv}{\vert\omega_n\vert}
\renewcommand{\v}[1]{{\bf #1}}
\begin{abstract}
The effects of electron-electron interactions on tunneling into the
bulk of a two dimensional electron system are studied near the integer
quantum Hall transitions.
Taking into account the dynamical screening of the interactions in the critical
conducting state, we show that the behavior of
the tunneling density of states (TDOS) is significantly altered 
at low energies from its noninteracting counterpart.
For the long-range Coulomb interaction, we demonstrate that the TDOS
vanishes linearly at the Fermi level according to a quantum Coulomb gap form,
$\nu(\omega)=C_Q\vov/e^4$, with $C_Q$ a nonuniversal
coefficient of a quantum mechanical origin.
In the case of  short-range or screened Coulomb interactions,
the TDOS is found to follow a power-law
$\vert\omega\vert^\alpha $, with 
$\alpha$ proportional to the bare interaction strength.
Since short-range interactions are known to be 
irrelevant perturbations at the non-interacting critical point, 
we predict that upon scaling, the power-law is smeared, leading to a
finite zero bias TDOS, $\nu(\omega)/\nu (0)=1+(\vov/\omega_0)^\gamma$
where $\gamma$ is a universal exponent
determined by the scaling dimension of short-ranged interactions.
We also consider the case of quasi-1D samples with edges, i.e. the
long Hall bar geometry, and find that the TDOS becomes dependent of the Hall 
conductance due to an altered boundary condition for diffusion. 
For short-range interactions, the 
TDOS of a quasi-1D strip with edges is linear near 
the Fermi level with a slope inversely proportional to $\rho_{xx}$ in 
the perturbative limit. These results are in qualitative agreement with
the findings of bulk tunneling experiments.
We discuss recent developments in understanding
the role played by electron-electron interactions at the 
integer quantum Hall transitions and the implications of these results on
dynamical scaling of the transition width. 
We argue that for long-range
Coulomb interactions, the existence of the quantum Coulomb gap in the quantum
critical regime of the transition gives rise to the observed
dynamical exponent $z=1$.

\end{abstract}
\pacs{PACS numbers: 73.40.Hm, 73.50.Jt,05.30.-d.}
\section{Introduction}

\subsection{Integer quantum Hall transitions and inadequacies of the 
noninteracting electron theory}

The physics of disorder and interaction in strong magnetic fields is
central to our understanding of the low-temperature, quantum mechanical
behaviors of novel electronic materials.
One of the most important physical phenomena under such settings is
the quantum Hall effect (QHE). \cite{von,tsui}
The QHE refers to the low-temperature
magnetotransport properties of high mobility two dimensional
electron systems (2DES) in a strong transverse magnetic field.
\cite{von,tsui,books}
The main part of the phenomenology can be summarized by: (1) The
existence of stable phases of matter, {\it i.e.} the quantum Hall states,
with vanishing dissipation and integer or fractional quantized Hall 
conductances; 
(2) The existence of continuous, zero temperature phase transitions
between the quantum Hall states, which is often referred to as the
the quantum Hall (plateau) transitions.
The basic physics in (1) for the spin-polarized incompressible quantum Hall
states and their low energy excitations are well 
understood.\cite{books,qhenobelrmp} In contrast, (2) is yet an unresolved 
problem, which is the subject of this work.

In a nutshell, (2) is a metal-insulator transition problem of 
the Anderson-Mott type in a 2D disordered system with strong time-reversal 
symmetry breaking. These transitions are generally believed to be prime 
examples of continuous quantum phase transitions, i.e. examples of 
quantum critical phenomena.\cite{sondhirmp,subir} 
Although there are reasons to suspect that the critical phenomena is
universal for both the integer and the fractional transitions,
\cite{engel90,jain,klz,lwk,huck3} we focus here on the
integer quantum Hall transitions (IQHT) in samples with sufficiently
strong disorder that fractional quantum Hall states do not intervene.
In this case, the transitions are directly between adjacent integer quantized 
Hall plateaus. The experimental data, reviewed in Ref.~\cite{sondhirmp},
can be summarized as follows:
(a) On either side of the transition
the Hall conductivity $\sigma_{xy}$ is quantized 
and the dissipative conductivity
has the limit $\sigma_{xx}\to0$ at zero temperature; 
(b) At the transition, $\sigma_{xy}$ is 
unquantized and $\sigma_{xx}$ remains finite at zero temperature, 
so that the disordered quantum critical state is conducting.
Thus the quantum phase transition is an unusual insulator to
insulator transition with no intervening metallic phase, only the
critical point itself has a finite conductance.

In an experimental situation, the divergent length, i.e. the critical
singularity is cut off by the presence of a finite
length scale, giving rise to a finite transition width within which
$\sigma_{xy}$ deviates from the quantized values and $\sigma_{xx}$ is
nonzero. The transition width, denoted as $\delta^*$ follows the scaling 
form,
\be
{\delta^*\over\delta_0} \sim {\rm min}\left[\left(\frac{L_0}{L}\right)
^{1/\nu_{\rm loc}},
\left(\frac{T}{T_0}\right)^{1/\zt\nu_{\rm loc}},
\left(\frac{\omega}{\omega_0}\right
)^{1/z_\omega\nu_{\rm loc}}\right],
\label{dbscaling} 
\ee
where $L$, $T$ and $\omega$ are the finite system size, temperature and 
measurement frequency in a specific experimental situation, 
and $\Delta_0$, $L_0$, $T_0$ and $\omega_0$ are microscopic scales. 
The various exponents in  Eq.~(\ref{dbscaling}) have the usual
meaning: $\nu_{\rm loc}$ is the static exponent of the single
divergent length scale, the localization length 
$\xi\sim \delta^{-\nu_{\rm loc}}$ where $\delta$ is
the distance to the quantum critical point;
$z_\omega$ is the dynamical exponent defining the length 
scale introduced by a finite frequency, $L_\omega\sim \omega^{-1/z_\omega}$;
and $\zt$ is the thermal exponent governing a temperature-dependent length
scale $L_T\sim T^{-1/\zt}$. In general, $\zt$ and $z_\omega$ can be 
independent exponents,\cite{wangetal} but $\zt=z_\omega$
for a generic quantum phase transition.\cite{sondhirmp,subir}
The three scaling regimes in Eq.~(\ref{dbscaling}) have all been
probed experimentally,\cite{hpwei,koch,hpnew,engel} as well as
the regime in which electric field strength sets the cut-off.\cite{chow}
The critical exponents extracted from the experiments can be summarized
as $\nu_{\rm loc}=2.3\pm 0.1$, $1/z_\omega\nu_{\rm loc}=0.41\pm 0.04$,
and $1/\zt\nu_{\rm loc}=0.42\pm 0.04$. Thus, we have  $\omega/T$-scaling with
$\zt=z_\omega=1$, which is in conformity with the
dynamical scaling description of a generic quantum phase transition.

The phase structure of the IQHT appears to be consistent with that of
the noninteracting theory of disordered 2D electrons in a strong 
magnetic field.\cite{pruiskenbook,pruiskenscaling} 
In a single-particle description,
all states are localized due to disorder, except for those at a single
critical energy $E_{c}$ near the 
center of each disorder-broadened Landau level.
The IQHT takes place when the Fermi level $E_{F}$ of the 2D 
electron system and one the discrete set of critical energy $E_{c}$
cross, i.e. when $\delta\equiv \vert E_{F}-E_{c}\vert$ approaches 
zero. Moreover, numerical calculations based on the noninteracting
theory give a localization length exponent 
$\nu_{\rm loc}\simeq 2.3$, which is remarkably
close to the experimentally value.

However, our understanding of the IQHT is far from complete. It has 
become increasingly clear that the noninteracting theory, reviewed
in Ref.~\cite{huck}, contradicts in several ways the experimental findings.
Three of these are in order.
(i) Recent experimental work has shown that the tunneling density
of states (TDOS) vanishes linearly at the Fermi level,
\cite{tunnelingexpt} in sharp contrast to the finite
density of states in the noninteracting theory.
(ii) It was pointed out recently that due to
the peculiar phase structure involving a set of 
extended states that has measure zero, the conductivity
$\sigma_{xx}$ in the noninteracting theory is rigorously zero in 
the limit of large sample size at all values of the magnetic
field, including the critical values, for any nonzero temperature.
\cite{wangetal} This is
in direct contradiction to the experimental observations.
(iii) The noninteracting theory does not offer a correct description
of the dynamical scaling behavior observed experimentally.
The dynamical exponent governing how the energy (temperature)
scale relates to the length scale for noninteracting electrons
is $z=d=2$, which {\it disagrees} with the experimentally obtained
values quoted above. In fact, the experimental findings of
$\omega/T$-scaling with $\zt=z_\omega=1$ is in conformity with
the dynamical scaling description of a generic quantum
phase transition in which Coulomb interaction is relevant
and scales to a finite value at the transition.\cite{fgg,lwinter} 

The failure of the noninteracting theory highlighted by (i)-(iii)
puts serious constraints on the ability of the free electron model
to explain the IQHT in real materials, and necessitates the investigation of
the effects of electronic interactions and their interplay
with disorder and localization.

\subsection{Recent theoretical developments on the effects of interactions
and the focus of this work}

A significant part of the recent theoretical studies on the role of
Coulomb interactions near the IQHT has centered around the three 
inter-connected issues (i)-(iii) raised in the 
above. \cite{yang1,yang2,lwinter,wangetal,thf,yang3,polyakov,qcgap}
There has also been formal approaches by 
Pruisken and collaborators that aim at extending the topological
nonlinear $\sigma$-model description of the noninteracting transition
to include Coulomb interactions.\cite{pinter}

The first quantitative
study of the one-particle density of state (DOS) in the presence of
Coulomb interactions was carried out by Yang and MacDonald.
Using a self-consistent Hartree-Fock (HF) approach, in which disorder
was treated exactly while the Coulomb interaction by HF approximation,
they found that the TDOS vanishes linearly at the Fermi level
at {\it all} filling factors in the lowest Landau
level, even at the critical energy.\cite{yang1} The linear Coulomb
gap behavior, especially at the critical energy, is in sharp
contrast to that expected of the noninteracting theory, see (i) above,
and is in qualitative agreement with 
experimental findings.\cite{tunnelingexpt}
In spite of the dramatic TDOS change due to Coulomb interactions,
however, Yang, MacDonald, and Huckestein found that the value
of the localization length exponent and the fractal dimension of the
critical eigenstate wave functions remain unchanged from the noninteracting
theory, as does the qualitative behavior of the conductivity.\cite{yang2}
It is important to emphasize a unique and important feature of the HF 
theory for the IQHT: the {\it noncritical} suppression of the single-particle 
DOS, i.e. it vanishes linearly at all filling fractions
regardless of the whether the system is at criticality or not. \cite{lwinter}

In order to understand the effects of Coulomb interactions
from the point of view of critical phenomena,
Lee and Wang carried out a stability analysis of the noninteracting fixed
point (NIFP), which governs the noninteracting transition,
by numerical calculations of the perturbative renormalization
group (RG) scaling dimensions for the interactions.\cite{lwinter}
They found that interactions of sufficiently short-range are perturbatively
irrelevant at the NIFP and scale to zero in the asymptotic
limit. The NIFP is therefore stable against such screened interactions, and as
a result, $\nu_{\rm loc}\simeq2.3$ and $z=2$. Wang, Fisher, Girvin, and
Chalker have shown that although short-range interactions are 
irrelevant in the RG sense, they generate a nonzero critical
value for the dissipative conductance, thus remove the pathology (ii) 
of the noninteracting theory, and control the temperature-scaling 
behavior of $\sigma_{xx}$.\cite{wangetal}
They showed that in the presence of irrelevant interactions,
the scaling theory for transport properties becomes
unconventional, $\omega/T$-scaling breaks down
and a third independent critical exponent, the
thermal exponent $\zt$ in Eq.~(\ref{dbscaling}), emerges. The value
of $\zt$ is set by the scaling dimension $-\alpha<0$ of the interaction
strength through the finite temperature dephasing time in the critical
regime, $\tau_\phi\sim T^{-p}$ where $p=1+2\alpha/z$, leading to
$\zt=2z/(z+2\alpha)$. They argued that quantum critical scaling
behavior of this kind may be a generic feature
of finite temperature transport near quantum critical points, when
interactions are (dangerously) irrelevant. \cite{wangetal}

In contrast to short-range, model interactions, true long-range Coulomb
interactions are, on the other hand, found to be relevant perturbations
at the NIFP, making the latter unstable.\cite{lwinter} Hence
the true critical point must be interacting, corresponding
to an interacting fixed point (IFP)  having a finite interaction strength.
This is consistent with the fact that the experimentally extracted
dynamical exponents $\zt= z_\omega=1$, which are typical of 
the charge dynamics at quantum criticality controlled by Coulomb
interactions.\cite{fgg} 
However, Lee and Wang proposed that the fixed point in the theory where
Coulomb interaction is treated via the HF approximation may 
in fact be stable.\cite{notehf}
They introduced the concept of a HF fixed point (HFFP) and argued that
it is the simplest possible interacting fixed point of the IQHT.
Correlation effects are found to be marginal perturbations at the 
HFFP due to the linear Coulomb gap in the HF theory that
degrades of the RG dimensions of the residual interactions.
They conjectured that a change in the dynamical exponent ($z$) with
no change in the static one ($\nu_{\rm loc}$) can be due to the
noncritical linear suppression of the single-particle (tunneling) DOS
induced by Coulomb interactions. The HF theory, in particular the
HFFP of the plateau transition, presents itself as a concrete example.
There are two important issues that must be resolved before 
this conjecture can be further substantiated. 

First, the theory of Coulomb gap was derived largely on the basis
of classical physics.\cite{ccgap} It applies directly to electronic
systems with Fermi energy lying in an excitation gap such
as semiconductors and insulators. Therefore it may not be completely
surprising that a 2D Coulomb gap DOS exists away from the transition
regime where the electronic states are strongly localized
and where the transport is dominated by variable range hopping
in the presence of a 2D Coulomb gap \cite{boris}.
What is remarkably surprising is that the linearly vanishing
Coulomb gap is found to pertain to the critical regime of the 
IQHT where the localization length
is enormously large and the conductivity finite.
This behavior is unprecedented and it is nature to ask whether
it is an artifact of the HF approximation that does not include the screening
of the exchange interactions.
Therefore, it is necessary to go beyond the HF theory in the
critical conducting regime and study the behavior of the TDOS 
when the screening of Coulomb interactions is taken into account.
In ordinary disordered metals in zero or weak magnetic fields,
the dynamical screening of the Coulomb interactions by the diffusive 
motion of the electrons is known to be very important.
\cite{aal,girvin,houghton}
It leads to {\it critical} corrections of the TDOS.\cite{bkrmp}
The natural question is whether the interplay between quantum diffusion
and Coulomb interaction at the IQHT leads to a linear Coulomb gap beyond
the HF theory.

This is the focus of the present work. In a recent Letter,\cite{qcgap}
we reported our findings that the quantum diffusive motion of the electrons, 
i.e. the diffusive dynamics, is too slow to effectively screen out the Coulomb
singularity in the dynamical case.  A nonperturbative resummation of the
the most singular corrections in the long time limit
to the TDOS gives rise to a linearly
vanishing TDOS for the critical conducting state. This behavior, termed
as the quantum Coulomb gap, can be thought as the quantum mechanical
analog of the classical Coulomb gap. It has  
a quantum origin and the slope of the gap
is nonuniversal in contrast to the classical case.
In this paper, we provide more physical and detailed theoretical
derivations of the quantum Coulomb gap. We also study the TDOS behaviors
for short range interactions, both outside the scaling regime where
a nonuniversal power-law TDOS is found, and in the scaling regime 
where the power law is smeared and a finite zero bias TDOS recovered 
in accordance with the observation that short range interactions are 
irrelevant perturbations in the RG sense. Interestingly, in the scaling
regime, the change in the TDOS, $\delta\nu(\ve)$ follows a power-law
with a universal exponent determined by the scaling dimension of
short-range interactions and the frequency exponent $z=2$ 
in this case. We also address in this paper the issue
of whether and how the bulk TDOS depends on the Hall conductance.
To this end, we study the case of quasi-1D samples with edges,
such as in the long Hall bar geometry, and find that the TDOS 
becomes dependent of the Hall conductance due to an altered
boundary condition for diffusion in a finite magnetic field.
It vanishes linearly at the Fermi level with a slope that
is inversely proportional to the magnetic field strength in the
perturbative regime, in good qualitative agreement with 
recent bulk tunneling experiments \cite{tunnelingexpt}.
These results will be summarized in the next subsection.

The second issue
has to do with the implications of the linear Coulomb gap on 
dynamical scaling. The linearly vanishing DOS in 2D means that the averaged
energy level spacing scales with the length of system according
to $\Delta_E\sim 1/L$, leading to a dynamical scaling exponent $z=1$.
However, one of the persistent mysteries remains, namely, it is not clear
that this is the dynamical exponent measured by the transport experiments.
The fact that quantum diffusion exists at the critical point of the
transition implies a frequency-dependent length scale
$L_\omega\sim ({dn/\over d \mu}\hbar\omega)^{-1/2}$ that is shorter than
the dynamical length scales derived from the single-particle sector.
Notice that the relevant DOS in $L_\omega$ is the thermodynamic
DOS or the compressibility $dn/d\mu$.\cite{palee}
Although it is somewhat unnecessary to associate a critical
exponent with diffusion, a value of $z_\omega=2$ is directly 
implied and it should govern the dynamics of diffusive transport 
in the asymptotic limit.
In a recent attempt to substantiate our previous conjecture made
in Ref.\cite{lwinter},
Huckestein and Backhaus \cite{thf} evaluated the density-density response 
function near the IQHT within a time dependent HF approximation
(TDHFA), in an effort to determine $z_\omega$ from
two-particle correlation functions.
Their analysis gives $z_\omega=1$, but under the compressibility
sum rule that relates $dn/d\mu$ to the static limit of the irreducible
density response function, it appears to have resulted
from using a linearly vanishing $dn/d\mu$ in
$L_\omega$. This result is at least counter-intuitive, since
$dn/d\mu$ is expected to be smooth and finite for a disordered system on
general grounds. Moreover, a finite compressibility is necessary for observing
the quantum Hall {\it transition} without the latter being interrupted by
incipient quantization plateaus. If $dn/d\mu$ were indeed vanishing, the
linear screening length would diverge and the screening properties
of the critical state would be similar to those of an insulator.
Recently, Yang, Wang, and MacDonald \cite{yang3} pointed out that the 
controversial result may be a consequence of not accounting for
the consistency of the exchange-local-fields and the disorder potential
in the TDHFA used. Analyzing the charge redistribution
following the insertion of an external test charge, they studied the
screening properties in the long wave
length limit of the self-consistent HF theory and find
that the thermodynamic DOS is finite despite of
the linearly vanishing tunneling DOS in the critical conducting
state. Therefore the question of whether or how the vanishing
single-particle DOS affects the dynamics of transport near the quantum
Hall critical point remains open. We will discuss this
issue in more detail in the last section of the paper.
%
%

The main part of this article is devoted to understanding how 
linearly-vanishing TDOS in 2D is likely so long as the conductivity is finite.
Similar analysis in the case of zero magnetic field has been carried 
out recently by Kopietz \cite{kop} in connection to the 2D $B=0$ 
metal-insulator transition. \cite{krav}
Our basic finding is that in the presence of disorder, 
Coulomb interaction is insufficiently screened by the quantum diffusive 
medium at finite frequencies. As a result, the single-particle DOS
in the extended regime comes to resemble that in the 
localized regime, i.e. exhibiting the linear Coulomb gap,
although the slope of gap is different due to a different mechanism.
After an understanding of the dynamics in the single-particle sector 
has been developed, we will turn to the important question of how the 
depletion of single-particle DOS relates to the larger issue of 
dynamical scaling near quantum phase transitions in disordered systems.

\subsection{Interplay between disorder and interaction}

At roughly the same time as the discovery of the integer quantum Hall 
effect, there were some remarkable developments in our understanding of
quantum transport such as localization and metal-insulator transitions.
\cite{gangof4} The weak localization theory was developed as a
perturbative approach to study the effects of disorder and
interactions. Early works by Altshuler and Aronov,
and Altshuler, Aronov, and Lee \cite{aal}
found several remarkable effects arising from the interplay of interaction and 
disorder: 1) the electron-electron scattering rate is enhanced due to 
the prolonged stay of electrons near one another;  2)  there is a 
correction to conductivity comparable to the localization effect 
caused by quantum interference;  3) and most dramatically,  
the TDOS is significantly altered from its noninteracting counter part
near the Fermi-energy. 

For ordinary disordered metals, perturbative diagrammatic calculations 
show that in 3D the weak localization correction to the TDOS, 
$\nu(\omega)$, is of the form $\dl \nu \sim \sqrt{\omega}$, a 
result largely confirmed by experiments in the early 1980s, \cite{aal}
where $\omega$ is measured from the Fermi energy.
In 2D, for long-range Coulomb interaction,
\be
\dl \nu= 
-\frac{1}{8\pi^2\hbar D}\ln (\omega \tau_0 )\ln(\omega\tau_1),
\label{deltanu}
\ee
indicating the possibility of a vanishing $\nu(\omega)$
near the Fermi-energy as $\omega\rightarrow 0$. 
In Eq.~(\ref{deltanu}), $D$ is the diffusion 
constant, $\tau_0$ is the elastic 
scattering time, and in terms of the inverse screening length
$\k=2\pi e^2 dn/d\mu$, $\tau_1$ is given by $1/\tau_1={\tau_0(D\kappa^2)^2}$.
Summation of all logarithmic terms are needed to find the limiting behavior.
This was done first by Finkel'stein in a field-theoretic
treatment of disorder and interaction. \cite{finkelshtein}
Defining the dimensionless conductivity {\it in units of } $e^2/\hbar$ via
the Einstein relation $\sigma={dn\over d\mu} D$, it was shown that
\be
\nu(\omega)=\nu_0e^{-{1\over8\pi^2\sigma}\ln(\omega\tau_0)\ln(\omega\tau_1)},
\label{resum}
\ee
which is valid {\it only} for $\sigma \gg ln(1/\omega\tau_0)$ such that
the weak localization correction to $\sigma$ can be ignored. The
conductivity is given by $\sigma=\nu_0D$ where $\nu_0$ is the finite
density of states in the self-consistent Born approximation (SCBA).
It should be emphasized that Eq.~(\ref{resum}) does not represent 
the asymptotic behavior of the TDOS at small bias $\omega$
where the conductivity is strongly renormalized and becomes 
itself scale-dependent. \cite{finkelshtein}

The behavior of $\nu(\omega)$ in metallic systems should be contrasted 
to the classical Coulomb gap behavior of the TDOS in disordered insulators.
Efros and Shklovskii \cite{ccgap} have shown that when the long-range
Coulomb interaction is unscreened, which is true in dielectric insulators,
the single particle DOS exhibits a universal Coulomb gap behavior,
\be
\nu_{\rm ES}(\omega)=\alpha_d \vert \omega\vert^{d-1}/e^{2(d-1)},
\label{classical}
\ee
where $d=3,2$ is the dimensionality and $\alpha_d$ is a dimensionless
constant of order unity. Thus in the insulating regime, one expects
a linearly vanishing Coulomb gap in 2D 
$\nu_{\rm ES}=\alpha_2\vert \omega\vert$.
Since the long-range $1/r$ Coulomb singularity is crucial in the
derivation of the classical Coulomb gap, it is only expected to be
valid in the strongly localized regime where the screening
of the interaction is weak and dielectric like.
In the quantum Hall effect, the latter corresponds
to the regions far away from the quantum Hall transitions.

Perhaps less well-known is that 
the same double-logarithmic correction
to the TDOS as given in Eq.~(\ref{deltanu})
was later derived {\it in the presence of a strong magnetic field}
by Girvin, Jonson, and Lee, \cite{girvin} and by
Houghton, Senna, and Ying. \cite{houghton}  Diffusion in a strong magnetic
field comes from the ``skipping'' of the semi-classical cyclotron orbits
caused by impurity scattering \cite{ando}. 
In SCBA,
the diffusion constant in Eq.~(\ref{deltanu}) is given by
$\dxx={1\over2}r_c^2 \tau_0^{-1}$, 
where the cyclotron radius $r_c=(2N+1)^{1/2}\ell_B$ 
with $\ell_B$ the magnetic length and 
$N$ the Landau-level index.  Note that in this case,
$\dxx$ is proportional to the field dependent scattering rate  
$1/\tau_0(B)\simeq[\omega_c/\tau_0(B=0)]^{1/2}$. 
In this work, we will derive the analog of Eq.~(\ref{resum})
in strong magnetic fields by a nonperturbative resummation of the 
double-log divergences. Since the critical conductance
is finite and scale-invariant at the IQHT,
it is possible for us to derive the
true asymptotic behavior of the TDOS in the low bias limit.
We show that, in the presence of disorder, Coulomb interaction is 
insufficiently screened by the 2D quantum diffusive medium at finite 
frequencies. As a result, the TDOS exhibits a linearly vanishing
quantum Coulomb gap behavior.

\subsection{Main results --- quantum Coulomb gap in TDOS}

The main results can be explained physically in a simple 
semi-classical picture.
The electron-electron interaction $v(\br-\brp)$ can be viewed as 
being mediated by a fluctuating potential field $\Phi(\br)$ 
with a distribution 
$P(\Phi) \sim \textstyle{e^{-\frac{1}{2}\Phi(\br)v^{-1}(\br-
\brp)\Phi(\brp)}}$. At the crudest level, neglecting all dynamical 
effects, the presence of a potential field directly changes the 
energy levels of individual electrons and contributes a phase delay
$e^{-i\int_0^{\tau} d\tau^\prime\Phi(\br_{cl}(\tau^\prime ))}$ to 
the single-electron propagator, where $\br_{cl}(\tau^\prime )$ is 
the classical trajectory. Such a semi-classical phase-approximation has been 
used recently in the context of composite fermions coupled to a 
fluctuating gauge field to study {\it edge tunneling}. \cite{KimWen,shytov}
Averaging the $\Phi$-field over different trajectories as well as the  
random potentials, we obtain the averaged phase lapse 
\be
e^{-W(\tau)}=\langle e^{-i\int_0^{\tau} d\tau^\prime \Phi(\br_{cl}
(\tau^\prime))} \rangle,
\label{phaselap}
\ee
during a time interval $(0,\tau)$. This phase delay can be viewed as a
Debye-Waller factor \cite{finkelshtein} for the impurity averaged
single-particle Green's function at
a (tunneling) site $\v r$,
$G(\tau)=\langle \psi(\v r,\tau)\psi^*(r,0)\rangle$,
\be
G(\tau)\simeq G_0(\tau)e^{-W(\tau)},
\label{debyewaller}
\ee
where $G_0(\tau)\sim 1/\tau$ is the counter-part of $G(\tau)$ in 
the absence of interactions. Notice that $G(\tau)$ no longer
depends on the coordinate ${\bf r}$ after impurity averaging.
The TDOS is given by
\be
\nu(\omega)=-{1\over\pi}{\rm Im}\int d \tau e^{i\omega_n\tau} G(\tau)
\vert_{i\omega_n\to\omega+i\eta}.
\label{tdos}
\ee
Two factors, both resulting 
from the diffusive nature of the electron motion in the presence of disorder,
lead to the divergence of the phase delay $W(\tau)$ at large $\tau$
and subsequently the vanishing of the single-particle DOS at the Fermi-energy: 
1) electrons stay longer in the vicinity of one another at each encounter, 
2) the Coulomb potential is not completely screened at finite time. 
Since we address the effects of interactions only up to a phase-delay, 
this part of the physics presumably can be set aside 
from the rest by performing a $U(1)$-rotation.
\cite{finkelshtein,qcgap,alex}

The specific form of the pseudogap in the TDOS depends on the type of the
interaction and on the scaling behavior of the interaction strength.
We shall consider both long-range and short-range, screened Coulomb
interactions.

\subsubsection{ Long range Coulomb interaction}

In the case of an long-range Coulomb potential, the phase delay diverges 
at long times as $W(\tau)\sim\ln(\tau/\tau_0)\ln(\tau/\tau_1)$. 
We will show in detail that this double-log divergence renders
the $\tau$-integral over $e^{-W(\tau)}$ convergent, whereby enabling
an expansion in $\omega$ for $\nu(\omega)$.
In the asymptotic low frequency limit, this leads
to a linearly vanishing TDOS at low temperatures
\be
\nu(\omega)= C_Q\hbar\vov/e^4 
\label{Coulomb}.
\ee 
We shall refer to Eq.~(\ref{Coulomb}) as the  
2D {\it quantum Coulomb gap behavior}. In contrast to the 2D classical
Coulomb gap behavior given in Eq.~(\ref{classical}),
the coefficient $C_Q$ in the quantum Coulomb gap is not a
universal number, but rather a quantity of quantum mechanical origin. 
It depends on the microscopic details of the 
sample such as the mobility. 
For large $\omega$, $\nu(\omega)$ crosses over to the  
perturbative diagrammatic result in strong magnetic fields.
\cite{girvin,houghton}

Since Coulomb interaction is a relevant perturbation at the
NIFP, the true transition must be governed by an interacting
fixed point (IFP) where the Coulomb interaction strength is finite.
Thus we expect that the quantum Coulomb gap to be the true
asymptotic behavior of TDOS at the integer quantum Hall transitions.
It is interesting to note that the Coulomb gap TDOS that we obtained 
for the critical conducting state at the IQHT
is qualitatively different from those obtained in the clean case 
\cite{he} and in a weak magnetic field. \cite{aleiner}

\subsubsection{Short range interactions --- prescaling regime}

For simplicity, we consider the case of a $\delta$-function 
interaction potential
$v(\br-\brp)=u\dl(\br-\brp)$ as a prototype short range
interacting potential. Outside the scaling regime, the scaling behavior
of the interaction strength $u$ can be ignored, i.e. $u$ can be
treated as a constant or equivalently as a marginal perturbation.
In this case, dynamical screening of the interaction
leads to a weaker, single-logarithmic divergence in the phase delay
$W(\tau)\sim \ln(\tau/\tau_0)$. 
The integral of the Debye-Waller factor
is no longer convergent, such that a power series expansion in $\omega$
becomes singular. This is similar the situation encountered in the
X-ray edge problem.\cite{mahan}  We find
a pseudogap in the TDOS that takes the form of a power-law,
\be
\nu(\omega)\simeq\nu_0 \vert\omega\tau_0\vert^{\al},
\label{short-ranged}
\ee
where the exponent $\al$ is nonuniversal and depends on the interaction
strength.
It is well known that transport at the quantum Hall transition 
in the noninteracting theory exhibits anomalous diffusion,\cite{cd}
i.e. the diffusion constant $D=D(q^2/\omega)\sim D_0(q^2\omega)^{\eta/2}$
when $D_0q^2>\omega$, where $\eta$ is a critical exponent related to
the multifractal dimension $D_2=2-\eta$. We will show that taking into 
account the anomalous diffusion, which has no effect in the Coulomb case, 
only leads to a weak $\eta$-dependence in the exponent $\alpha$ in 
Eq.~(\ref{short-ranged}).

\subsubsection{Short range interactions --- scaling regime}

Because short range interactions are irrelevant perturbations at the
NIFP,\cite{lwinter} the strength of the effective interaction $u$ 
must scale to zero in the scaling regime according to 
$u_{\rm eff}\sim u\omega^{x_+/z}$, where $-x_+\simeq-0.64$ is the dimension
of the interaction and $z=2$ is the dynamic exponent at the stable NIFP.
This makes the phase delay $W(\tau)$
converge in the large-$\tau$ limit. As a result, the power-law
decay in Eq.~(\ref{short-ranged}) is smeared, resulting in a
finite zero-bias TDOS,
\be
\nu(\omega)=\nu(0)\left[1+\left({\vov\over\omega_0}\right)^\gamma\right],
\label{nusr} 
\ee 
where $\omega_0$ is a frequency scale and
$\gamma=x_+/z\simeq0.32$ is a universal exponent.
This result leads to several interesting predictions: (a) For short-ranged 
interactions, the TDOS is finite and nonuniversal at zero bias.
(b) It can be shown that  $\nu(0)\ll\nu_0$ if the bare interaction strength
is strong, so short-range 
interactions irrelevant in the RG sense can still lead 
to strong density of states suppression at low bias. (c) 
Eq.~(\ref{nusr}) shows that the TDOS increases 
with $\omega$ according to a {\it universal} power law 
with an initial cusp singularity for our value of $\gamma$. 
These predictions can, in principle, be tested experimentally since
the Coulomb interaction can be made short range by placing
a metallic screening gate (ground plane) nearby.

\subsubsection{Quasi-1D samples with edges}

We also study whether and under what condition
the bulk TDOS depends on the Hall conductance. 
That $\nu(\omega)$ in
Eqs.~(\ref{Coulomb},\ref{short-ranged},\ref{nusr}) 
does not depend on $\sigma_{xy}$
is a direct consequence of the fact that transverse force does not
affect the charge spreading in the bulk of the sample. Thus any direct
$\sigma_{xy}$-dependence in tunneling must come from contributions 
at the boundary. It has been shown by Khmel'nitskii and Yosefin, \cite{KY}
and by Xiong, Read and Stone \cite{XRS} that in the presence of edges, 
the Hall conductance enters measurable quantities even in the perturbative 
limit. More recently, Shytov, Levitov, and Halperin studied the
problem of {\it edge}
tunneling into the fractional quantum Hall state where the Hall conductivity
dependence of the I-V characteristics also arises from the
boundary condition at the tunneling edge.\cite{shytov}

We considered a quasi-1D sample with its length $L$ much 
greater than its width $W$ and with two reflecting edges. This condition
is realized experimentally in the long Hall bar geometry.
Edge effect becomes prominent in such limit since the boundary condition 
effectively changes 
the diffusion constant from $D$ in 2D to $D_{1D}=D(1+\gamma_H^2)$ 
where $\gamma_H=\frac{\sigma_{xy}}{\sigma_{xx}}$ is the Hall ratio.
For the case of $\del$-function interaction, we find that the asymptotic TDOS
becomes linear in an infinite strip with edges:
\be
\nu(\omega)= s\vov.
\label{1Dedge}
\ee 
The slope of the linear gap is proportional to 
the inverse of the dissipative resistivity $s\sim\rho_{xx}^{-1}$. 
In the high field limit where energy levels form Landau bands 
($\omega_c\tau_0\gg 1$), the bare value of $\sigma_{xx}$ in SCBA 
at the center of the 
Landau levels is proportional to the Landau index $N$ while the bare
Hall ratio is of order $1$. 
The above result then implies that, in the perturbative regime where
the localization correction to the conductivity is much smaller
than the SCBA conductivity, the slope of linear density of states
is proportional to $N$ or $1/B$.
Interestingly such a dependence was indeed observed in the high field 
bulk tunneling experiments.\cite{tunnelingexpt}
It remains to be seen whether the samples used in certain experimental
setup can be qualifies as being quasi-1D with edges and the presence 
of ground planes indeed make the Coulomb
interaction short-range.  

\subsection{Organization of the paper}

In Section II, we revisit the role of electronic
interactions at the NIFP of the quantum Hall
transition, and cast the results of Lee and Wang \cite{lwinter} on
the RG dimensions of interactions in an
the analysis of the level spacing in the Hartree-Fock theory.
In doing so, we make connections to the more conventional scaling
theory of localization a la Wegner \cite{wegner} and motivate
the study of the single particle DOS.
We then proceed, in Section III, to formulate the effective field theory 
and the semi-classical phase approximation, and to derive the 
Debye-Waller factor in Eq.~(\ref{debyewaller}). The bulk TDOS in
2D is calculated in Section IV for various forms of interactions
in the perturbative and the scaling regimes. The results summarized
in Eqs ~(\ref{Coulomb}),(\ref{short-ranged}),(\ref{nusr}) 
are derived in this section.
The effect of the anomalous diffusion at the IQHT is also studied.
Section V is devoted to the derivation of the TDOS in quasi-1D samples
in the presence of edges, Eq.~(\ref{1Dedge}). 
Summary and discussions on the connection between
the single-particle DOS and the dynamical scaling of the transition width
are given in Section VI. We argue that
the existence of the quantum Coulomb gap in the quantum critical
regime of the transition gives rise, through the
interplay between quasiparticle decay rate and level spacing,
to the experimentally observed dynamical exponent $z=1$.

\section{Role of Interactions --- Hartree-Fock analysis of Level spacing}

The Hamiltonian of interest 
describes interacting electrons moving in a two-dimensional
random potential in the presence of a magnetic field:
\be
H=\sum_i\left[{1\over2m}\left({\v
p}_i+{e\over c}{\v A}\right)^2 +V(\v r_i)\right] +H_{\rm int},
\label{h}
\ee
where ${\v A}$ is the external vector potential producing the strong
transverse magnetic field, $V(\v r)$ is the
one-body impurity potential. The interacting part of the Hamiltonian
$H_{\rm int}$ is given by 
is the two-body interaction potential,
\bea
H_{\rm int}&=& {1\over2} \sum_{i\neq j}v({\v r}_i -{\v r}_j), 
\label{hint} \\
v({\v r}_i-{\v r}_j)&=&{u\over\vert {\v r}_i-{\v r}_j \vert^p}.
\label{vint}
\eea
Here $u$ and
$p$ control the strength and the range of the
interaction. The Coulomb potential corresponds 
to $u=e^2/\epsilon$ and $p=1$, and a short distance cutoff 
can be introduced for the case of $p\ge4$.

The noninteracting fixed point of the IQHT is obtained by setting $u=0$.
The effects of interactions can be studied in the framework of
critical phenomena by analyzing the stability of the NIFP.
Imagine starting with a system at the NIFP, and adiabatically
switching on the interaction $u$. One can ask whether $u$ is a
relevant or irrelevant perturbation in the renormalization group (RG)
sense by calculating the RG scaling dimension of $u$. This has been
done by Lee and Wang.\cite{lwinter} They analyzed the most singular
contributions to the disordered averaged free energy functional.
Here we present an alternative treatment \cite{huck2}
from the point of view of the single-particle DOS
and cast the result of Lee and Wang in terms of the interaction induced
corrections to the single-particle level spacing.

For this purpose, it is convenient to use the exact eigenstates approach
\cite{elihu} at criticality. Consider a finite system of linear dimension
$L$, and two adjacent one-electron eigenstates
located on the two sides of the critical energy $E_c$
with energy $E_1< E_c$ and $E_2>E_c$ and a separation $\omega=E_1-E_2$.
In the noninteracting theory, the finite size scaling behavior
of the separation should follow that of the mean level spacing 
and scale with $L$ according to 
\be
\Delta_{12}^0= {1\over\nu_0L^d}\propto{1\over L^2},
\label{del0}
\ee
where $\nu_0$ is the noncritical DOS in the noninteracting theory.
Switching on the interaction $u$ causes a mixing of the critical 
eigenstates, which results in shifting of the single-particle energy levels. 
The level spacing becomes,
\be
\Delta_{12}=\Delta_{12}^0+\delta\Delta_{12},
\label{del}
\ee
where $\delta\Delta_{12}$ is the level shift of $E_2$
due to the mixing with level $E_1$. The question we would like
to ask is, in the limit $\omega\to0$ and the associated length scale
$L_\omega=(\nu_0\omega)^{1/d}\to\infty$, how the interaction correction to
the level spacing $\delta\Delta_{12}$ scales with $L$
as we approach the thermodynamic limit $L< L_\omega\to\infty$.
If it falls off faster than
the mean level spacing $\Delta_{12}^0\sim 1/L^2$ of the noninteracting system,
the level statistics will be determined by that of 
the noninteracting eigenstates 
and unaffected by the interaction in the thermodynamic limit.
Thus the interaction would be an irrelevant perturbation in the
RG sense. On the other hand, if $\delta\Delta_{12}$ falls off slower than
$1/L^2$, the level spacing will be dominated by the interaction
induced level shifts as an increasing number of noninteracting eigenstates
gets mixed by the interaction with increasing system size $L$.
The interaction would therefore be a relevant perturbation 
in this case, and the noninteracting fixed point would be unstable.

Let's calculate $\delta\Delta_{12}$ perturbatively, which is sufficient
for the perturbative RG analysis.
To first order in perturbation theory, we have
\be
\delta\Delta_{12}=P_1\langle 2\vert H_{\rm int}\vert 2\rangle,
\label{del12}
\ee
where $P_1$ is a projection operator that keeps exclusively the contribution
from state $\vert 1\rangle$ to the level shift of $\vert2\rangle$. 
The factorized
interaction $H_{\rm int}$ can be written as
\be
H_{\rm int}=\sum_\alpha \Sigma_\alpha^{HF} c_\alpha^\dagger c_\alpha.
\label{hinthf}
\ee
where $\Sigma^{\rm HF}$ is precisely the Hartree-Fock self energy 
correction to the single particle state $\vert 
\alpha\rangle=c_\alpha^\dagger \vert 0\rangle$,
\be
\Sigma_\alpha^{HF}=\sum_\beta^{\rm occ.}\int d^2 \v rd^2 \v r^\prime
\left[\vert\psi_\alpha(r)\vert^2\vert\psi_\beta(r^\prime)\vert^2
-\psi_\alpha(r)^*\psi_\alpha(r^\prime)\psi_\beta^*(r^\prime)
\psi_\beta(r)\right]v({\v r} -{\v r}^\prime).
\label{se}
\ee
Here the summation is over all occupied states.
To discuss the average energy shift, it is necessary to study
the disorder average of the self-energy at the fixed energy $E_2$,
\be
{\Sigma_2}={1\over\nu_0L^2}\sum_\alpha\delta(E_2-E_\alpha)
\Sigma_\alpha,
\label{se2}
\ee
where $\nu_0$ is the DOS per unit area for noninteracting electrons.
Taking the disorder average of Eq.~(\ref{del12}) and using 
Eqs.~(\ref{hinthf}), (\ref{se}), and (\ref{se2}), 
we obtain the averaged level shifts,
\bea
{\overline{\delta\Delta_{12}}}&=&
\left({1\over\nu_0L^2}\right)^2\int d^2 \v rd^2 \v r^\prime
{\overline{\sum_{\alpha\beta}
\left[\vert\psi_\alpha(r)\vert^2\vert\psi_\beta(r^\prime)\vert^2
-\psi_\alpha(r)^*\psi_\alpha(r^\prime)\psi_\beta^*(r^\prime)
\psi_\beta(r)\right]\delta(E_1-E_2-\omega)}}v({\v r} -{\v r}^\prime).
\nonumber \\
&=&\left({1\over\nu_0L^2}\right)^2\int d^2 \v rd^2 \v r^\prime \left[
O_A(\v r -\v r^\prime)-O_B(\v r -\v r^\prime)\right]
v({\v r} -{\v r}^\prime).
\label{delfinal}
\eea
In order calculate this quantity, we need to know the scaling behavior
of the impurity average products of four wave functions denoted by
$O_A$ and $O_B$ in Eq.~(\ref{delfinal}) in the limit $\omega\to0$. 
The latter are functions of $\v r-\v r^\prime$ since translation
symmetry is restored after impurity averaging. 

It is instructive to follow Wegner's symmetry decomposition
\cite{wegner,pruiskenip,bkip} and extract the orthogonal (eigen) scaling 
variables under the RG. Consider the most general four-field operator 
in the unitary ensemble,
\be
O_4=\sum_{\alpha\beta\gamma\delta} v_{\beta\delta}^{\alpha\gamma}
\ \psi_\alpha^*\psi_\beta\psi_\gamma^*\psi_\delta
\equiv\sum_{\alpha\beta\gamma\delta} v_{\beta\delta}^{\alpha\gamma}
\ O_{\alpha\gamma}^{\beta\delta}, 
\label{o4}
\ee
where $O_{\alpha\gamma}^{\beta\delta}=O_{\gamma\alpha}^{\delta\beta}$.
The coefficients $v$ obeys the traceless condition, corresponding to
the subtraction of vacuum expectations,
\be
\sum_i v_{i\delta}^{i\gamma}=\sum_j v_{\beta j}^{j\gamma}=0.
\label{traceless}
\ee
There are two irreducible representations for these operators,
a symmetric one and an anti symmetric one, under permutations of
indices of the rank two tenser. We can therefore decompose $O_4$ into
independent scaling operators,
\be
O_{\alpha\gamma}^{\beta\delta}={1\over2}O_+ + {1\over2}O_-, \qquad
O_\pm=O_{\alpha\gamma}^{\beta\delta}\pm O_{\alpha\gamma}^{\delta\beta}.
\label{opm}
\ee
The operators $O_+$ and $O_-$, having independent scaling dimensions $x_\pm$,
describe the eigen scaling directions of the four-field operators under the RG.
In the unitary universality class of the metal-insulator transition 
i.e. the cases of weak magnetic field and spin-flip scattering
by magnetic impurities, $x_\pm=\pm\sqrt{2\epsilon}$ has been derived
from perturbation theory in $2+\epsilon$ dimensions. \cite{pruiskenip}

At the IQHT, the RG dimensions of 
$O_\pm$ have been determined numerically by Lee and Wang,
\cite{lwhub,lwinter} from the leading scaling operators associated 
with the fusion products of four fermion operators that are antisymmetric and  
symmetric under permutations respectively. The scaling dimension of $O_-$ is
obtained from the product of two ``spin'' operators \cite{lwhub}
\be
x_-= x_{2s}=-0.60\pm.02,
\label{x-}
\ee
whereas that of $O_+$ is extracted from the leading
scaling operator fused by the product of two nearby
density operators \cite{lwinter}
\be
x_+= x_{2\rho} =0.65\pm.04.
\label{x+}
\ee
A general four-field operator involves contributions from both $O_-$ and 
$O_+$, but its leading scaling behavior will be dominated by that of $O_-$, 
since $O_-$ is much more relevant than $O_+$. A good example \cite{lwhub}
is the ensemble averaged inverse participation ratio introduced by Wegner
\cite{wegner} $P^{(2)}$. Expressed in terms of a four-field operator,
its scaling dimension, which is also known as the multifractal
dimension of the eigenstates, $D(2)$, is governed by $x_-$, i.e.
$D(2)=d+x_-=1.4\pm0.02$, indicating strong amplitude fluctuations
of the critical eigenstates at plateau transitions. Indeed, the
exponent $\eta$ used by Chalker and Daniel \cite{cd} to
describe the anomalous diffusion at IQHT is given by
$\eta=-x_-$. The scatter in the value of $x_-$ is most likely
due to uncertainties involved in different numerical approaches.
By the same token, one can show that the scaling behavior of the
ultrasonic attenuation, extensively studied in $2+\epsilon$ dimensions
near the conventional metal-insulator transitions \cite{kotcas},
is controlled by $O_-$ as well.

Now let's apply these results to the interaction induced level shifts
in Eq.~(\ref{delfinal}). We will show that the density-density
correlation is, in contrast to the inverse participation ratio and the 
anomalous diffusion coefficient, controlled by the symmetric operator 
$O_+$ with the scaling dimension $x_+$.
Individually, operators $O_A$ and $O_B$
contain contributions from both $O_-$ and $O_+$. Their leading scaling
behavior is therefore dictated by that of the operator $O_-$. We have
\be
O_A(\v r-\v r^\prime)\sim O_B(\v r -\v r^\prime)
\sim\left(\vert r-r^\prime\vert\over L\right)^{x_-},
\label{leading}
\ee
for $\vert r-r^\prime/L\ll L_\omega$. 
However, the combination of
$O_A-O_B$ has precisely the symmetry of the symmetric operator $O_+$. Thus,
\be
O_A(\v r-\v r^\prime)- O_B(\v r -\v r^\prime)
\sim \left(\vert r-r^\prime\vert\over L\right)^{x_+}.
\label{delscaling}
\ee
We see that the scaling
behavior of the interacting-induced level shift is determined by
the fusion product of two 
density operators in the symmetric representation.
The absence of pure powers of
$\vert r-r^\prime\vert$ in Eqs.~(\ref{leading}) and (\ref{delscaling})
comes from the fact that bi-linear field operators have dimensions
zero,\cite{lwhub,lwinter} consistent with the single-particle DOS
being non-critical at the noninteracting critical point.

Substituting Eq.~(\ref{delscaling}) into Eq.~(\ref{delfinal}) and carry
out the spatial integrals from a lattice cutoff $a$ to the system size $L$,
we obtain,
\be
{\overline{\delta\Delta_{12}}}={u\over L^p}\left[c_1+c_2
({a\over L})^{2+x_{+}-p}\right],
\label{levelfinal}
\ee
where $c_1$ and $c_2$ are nonuniversal constants. Notice that for
$p>2+x_{+}$, the integral depends on the lower cutoff
and the second term in Eq.~(\ref{levelfinal}) diverges as $a\to0$ which
must be absorbed into the renormalized interaction.
We are now ready to determine the relevance of the interactions by comparing
the scale dependence of ${\overline{\delta\Delta_{12}}}$ to the mean
level spacing of the noninteracting system for large $L$.
Define the scaling dimension of the interaction $u$ according to,
\be
x={d\over d\ln L}\left[
\log\left( {\overline{\delta\Delta_{12}}}\over
{\overline{\Delta_{12}^0}}\right)\right],
\label{defalpha}
\ee
we obtain using Eq.~(\ref{levelfinal}) and ${\overline{\Delta_{12}^0}}\sim
L^{-2}$
\be
x={\rm max}(2-p,-x_{+}).
\ee
\begin{figure}    
\center    
\centerline{\epsfysize=2.8in    
\epsfbox{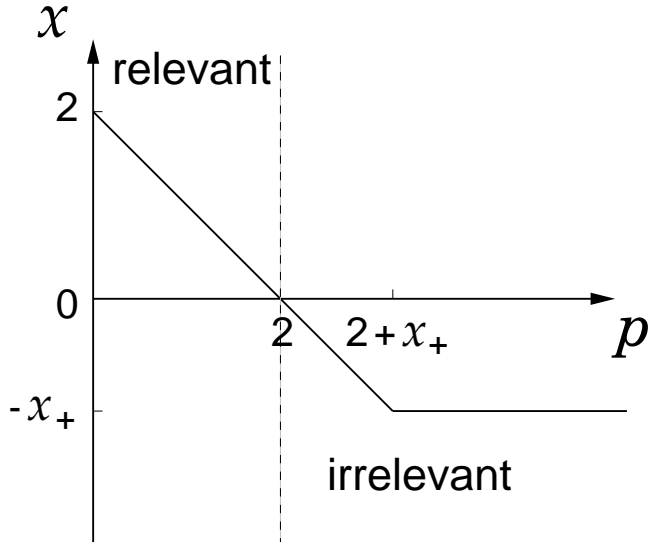}}    
\begin{minipage}[t]{8.1cm}     
\caption{The scaling dimension $x$ of the interaction strength $u$ 
in a $1/r^p$-potential as a function of $p$. The interaction is
relevant for $x>0$ and irrelevant for $x<0$ in the RG sense.
}   
\label{fig1}      
\end{minipage}     
\end{figure}    
The behavior of $x$ is shown in Fig.~1.
For $p<2$, the interactions acquire a RG scaling dimension $x=2-p>0$ 
and are relevant. We can refer to these type of interactions as long 
range interactions. \cite{lwinter} In this case, the interaction
induced level shift becomes much larger than the mean level spacing of the 
critical eigen states in the noninteracting theory for large system sizes $L$.
For the Coulomb interaction, $p=1$, $u$ has a
RG scaling dimension $x_{\rm Coul.}=1$ and is therefore a relevant 
perturbation. The resulting flow away from the NIFP will lead to an
interacting fixed point at which the effective interaction strength is finite. 
Presumably, the simplest version of the latter is the Hartree-Fock fixed 
point discussed in the previous section. At the level of the Hartree-Fock 
theory, Eq.~(\ref{levelfinal}) shows that the
level spacing is entirely dominated by the Coulomb interaction
induced level shift that scales as $L^{-1}$ in the thermodynamic limit, 
consistent with the linear Coulomb gap DOS found in numerical 
calculations.\cite{yang1}

On the other hand, for all values of $p>2$, we have $x<0$. The
interactions are irrelevant and can be refereed to as short range interactions. 
The dipole-dipole interaction, in particular, having $p=3$, belongs
to this class of interactions. 
The NIFP is therefore stable
against short range interactions. For screened Coulomb interaction
with $2<p<2+x_{+}$, the RG dimension of $u$ is $x=2-p$, while
for $p>2+x_+$, $x=-x_+$. In both cases the interaction scales
to zero at the transition in the asymptotic limit, although, it
controls the finite temperature behavior of the conductances.\cite{wangetal}
From Eq.~(\ref{levelfinal}), it is clear that the interaction induced
mixing between the critical eigenstates of the noninteracting theory
only leads to level shifts that are much smaller than the mean level
spacing in the thermodynamic limit. Thus, the
zero bias DOS must be finite in the asymptotic limit for short range
interactions. However, as we will demonstrate later 
in this paper, even in this case, the interactions lead to
remarkable properties of the TDOS in the pre-scaling 
regime, which may have important experimental consequences.

\section{Field theory framework and semi-classical approximation}

In order to include the screening
of the Coulomb interactions by the diffusive electrons in the calculation
of the TDOS beyond the Hartree-Fock theory, 
we will set up, in this section, the effective field theory and
the semiclassical approximation in order to derive the Debye-Waller factor
in Eq.~(\ref{debyewaller}).

\subsection{General formalism }

We consider the action for interacting electrons in a 
random potential and a magnetic field described by the Hamiltonian
in Eq.~(\ref{h}),
\be
S=\int_0^{\beta}d\tau d^2 \br{\cal  L},
\label{s}
\ee
where
\be
{\cal L}=\psi^*\left[\partial_{\tau} +H_0+V(\br)\right]\psi+\frac{1}{2}
\int d^2\br 
\psi^*(\br)\psi(\br)v(\br,\brp)\psi^*(\brp)\psi(\brp).
\label{l}
\ee
Here, once again $V(\br)$ is the random external potential, $v(\br-\brp)$ is 
the electron-electron interaction, and $H_0= 
\frac{1}{2m}(\partial_i+ieA_i)^2$, 
with ${\bf A}(\br)$ being the vector potential of a uniform 
external magnetic field 
perpendicular to the 2D plane. In Eq.~(\ref{l}), $\psi$ and $\psi^*$ are
independent Grassmann fields.
The electron single-particle Green's function is defined as:
\be
G(\br, \br,\tau)={\cal Z}^{-1}\int D[\psi^*] D[\psi]
\psi(\br,0)\psi^*(\br,\tau) e^{-S},
\label{Green's function}
\ee
where ${\cal Z}$ is the partition function 
expressed in terms of the imaginary time path integral,
\be
{\cal Z}=\int D[\psi^*] D[\psi]  e^{-S}.
\ee
The interaction between the electrons can be viewed as being 
mediated by a fluctuating scalar field $\Phi$ in the Coulomb gauge, 
for one can always rewrite the 
interaction term in the following way:
\be
e^{-\frac{1}{2}\int d^2\br \int d^2\br 
\psi^*(\br)\psi(\br)v(\br,\brp)\psi^*(\brp)\psi(\brp 
)}=\int D[\Phi] e^{{
i\int d^2\br \psi^*(\br) \Phi(\br) \psi(\br) -\frac{1}{2}\int 
d^2\br \int d^2 \brp 
\Phi(\br)v^{-1}(\br-\brp)\Phi(\brp)}}.
\ee
To perform the averaging over random potential, we use the replica trick,
calculate ${\cal Z}^n$, where $n$ is the number of replicas, and take the 
limit $n\rightarrow 0$ at the end. The ensemble averaged single-particle
Green's function can be obtained according to,
\bea
\langle G(\tau) \rangle&=&\lim_{n\rightarrow 0}\int D[\Phi]\int D[V]P[V] 
\int D[\psi^*] D[\psi] 
\psi_{\al_1}(\br,0)\psi_{\al_1}^*(\br,\tau)\non\\
&& \exp\left\{-\int_0^{\beta}d\tau \sum_{\al=0}^{n} \left[\int d^2\br
\psi^*_{\al}(\br,\tau)[\partial_{\tau}+H_0+V-
i\Phi_{\al}(\br,\tau)]\psi_{\al}(\br,\tau)
\right.\right.\non\\
&&+\left.\left.\int d\br \int d \br^\prime  \Phi_{\al}(\br,\tau)v^{-1}(\br-
\brp)\Phi_{\al}(\brp,\tau)\right]\right\}.
\label{avgG}
\eea
In the above equation, $P[V]$ is the distribution of the random potential 
which is taken to be
Gaussian white noise: $P[V]\sim e^{-\frac{1}{2g} V^2(\br)}$
for the short-range correlated impurities,
$\al$ is the replica index, and $\al_1$ represents an arbitrary
replica channel.
As in the usual treatment of disordered systems, integrating over $V(\br)$ in
Eq.~(\ref{avgG}) introduces a 4-point interaction term 
in the action that is local
in space but nonlocal in time,
$\frac{g}{2}\textstyle{\sum_{\al}\sum_{\al^\prime }}\int d^2\br 
\int_0^{\beta} d\tau\int_0^{\beta} d\tau^\prime 
|\psi_{\al}(\br,\tau)|^2|\psi_{\al^\prime }(\br,\tau^\prime )|^2$.
Th latter is usually decoupled by
introducing an auxiliary field $Q_{\al,\al^\prime }(\tau,\tau^\prime)$ 
by way of the Hubbard-Stratonovic transformation:
\bea
&&e^{{\frac{g}{2} \int d^2\br \int_0^{\beta} d\tau\int_0^{\beta} 
d\tau^\prime  
|\psi_{\al}(\br,\tau)|^2|\psi_{\al^\prime }(\br,\tau^\prime )|^2}}\non\\
&&=\int D[Q] e^{{-\frac{1}{2g}\int d^2\br 
\int_0^{\beta} d\tau\int_0^{\beta} d\tau^\prime  
Q_{\al\al^\prime }(\tau,\tau^\prime )Q_{\al^\prime \al}(\tau^\prime ,\tau) 
+i \int d^2\br \int_0^{\beta} 
d\tau\int_0^{\beta} d\tau^\prime  
\psi^*_{\al}(\tau)Q_{\al\al^\prime }(\tau,\tau^\prime )
\psi_{\al^\prime }(\tau^\prime )}}.
\eea
This quench-averaging process leads to the following replicated action,
\bea
S({\psi^*}, \psi, Q, \Phi )&=&\int d^2\br \int_0^{\beta} d\tau
\biggl[\int_0^{\beta} d\tau^\prime 
\sum_{\al\al^\prime}\bigl\{\psi^{*}_{\al}(\br,\tau)\left\{[\partial_{\tau}+
H_0-\Phi_{\al}(\tau)]\delta(\tau-\tau^\prime )\delta_{\al\al^\prime }
\right.\nonumber \\
&-&\left.
iQ_{\al,\al^\prime }(\tau,\tau^\prime )\right\}\psi_{\al^\prime }
(\br,\tau^\prime )
+\frac{1}{2g}Q_{\al\al^\prime }(\tau,\tau^\prime )Q_{\al^\prime \al}
(\tau^\prime \tau)\bigr\}
\nonumber \\
&+&\frac{1}{2}\sum_\al\int d^2\brp 
\Phi_{\al}(\br,\tau)v^{-1}(\br,\brp)\Phi_{\al}(\brp,\tau)\biggr].
\label{s2}
\eea
The impurity-averaged Green's function in Eq.~(\ref{avgG}) becomes,
\be
\langle G(\tau)\rangle=\lim_{n\to0}\int D[\Phi]\int D[Q]
\int D[\psi^*] D[\psi]
\psi_{\al_1}(\br,0)\psi_{\al_1 }^*(\br,\tau)
e^{{-S(\psi^*,\psi,Q,\Phi)}}.
\label{avgG2}
\ee
The rest of this section proceeds as follows: 
1) We integrate out the $Q$-field as well as the $\psi$-field and derive an 
effective action in terms of the $\Phi$-field,
$S_{\rm eff}(\Phi)$. The Green's function $\langle G(\br,\brp;\tau)\rangle$ 
can be expressed in terms of the averaged non-interacting electron Green's 
function  in the presence of the fluctuating potential $\Phi$, 
which we denote by $\ov{ G}(\Phi)$, weighted by $e^{-S_{\rm eff}(\Phi)}$. 
The effective action $S_{\rm eff}$ can be obtained systematically
in a power series of $\Phi$ and $1/{\sigma}_{xx}$. 2) By appealing  
to the semi-classical approximation for the slowly-varying part of the 
$\Phi$-field, we argue that $\ov{G}(\Phi)\approx  \ov{G}(0) 
\exp(-i\int_0^{\beta} d\tau\int d^2\br \Phi \rho) $, where $\rho(\br)$ is
the diffusion propagator. 3) Keeping in $S_{\rm eff}$ up to quadratic
terms in $\Phi^2$ and integrating out the $\Phi$ field,  
we arrive at $\langle G\rangle \sim \ov{G}(0) e^{-W(\tau)}$.  

\subsection{The effective action}
Let us define the effective action by formally integrate out the $Q$ 
and $\psi$-field,
\be
{\cal Z}[{\Phi}]=e^{-S_{\rm eff}(\Phi)}=
\int D[Q]\int D[{\psi^*}] D[\psi] e^{-S({\psi^*},\psi,
Q,\Phi)}.
\label{zpsi}
\ee
Carrying out explicitly the functional integral over the $\psi$-field,
we obtain
\be
{\cal Z}[\Phi]=e^{-S_{\rm eff}(\Phi)}=\int D[Q] e^{-S_Q(\Phi)-S_2^{(0)}(\Phi)},
\label{z2}
\ee
where, 
\bea
S_Q(\Phi)&=&-\frac{1}{2g}\Tr Q^2 +\Tr\ln[i\omega -H_0+i\Phi +iQ],
\label{sq} \\
S_2^{(0)}(\Phi)&=&\frac{1}{2}
\Tr \Phi(\br)v^{-1}(\br,\brp)\Phi(\brp).
\label{s20}
\eea
In Eq.~(\ref{sq}), $\omega$ is the fermion Matsubara frequency and
``Tr''  stands for the trace over the replica, spatial, and Matsubara indices.
Expanding $\Tr\ln[i\omega -H_0+i\Phi +iQ]$ in power series of $\Phi$,
we get 
\bea
S_Q&=&
-\frac{1}{2g}\Tr Q^2 +\Tr\ln[i\omega -H_0 +iQ]+ \Tr\left(\frac{1}{i\omega-
H_0+iQ}i\Phi\right)\non\\
&&-\frac{1}{2}\Tr\left(\frac{1}{i\omega-H_0+iQ}i\Phi\frac{1}{i\omega-
H_0+iQ}i\Phi\right)+{\cal O}(\Phi^3).
\label{sqexpand}
\eea
We group the terms in $S_Q(\Phi)$ which are zeroth order in $\Phi$
as $S_{\sigma}$, 
\be
S_{\sigma}=-\frac{1}{2g}\Tr Q^2 +\Tr\ln[i\omega -H_0 +iQ],
\label{ssigma}
\ee
the terms linear in $\Phi$ as $\Delta S_1(\Phi)$,
\be
\Delta S_1= \Tr\left[\frac{1}{i\omega-H_0+iQ}i\Phi\right],
\label{ds1} 
\ee
and the $\Phi^2$-term as $\Delta S_2(\Phi)$,
\be
\Delta S_2= -\frac{1}{2}\Tr\left[\frac{1}{i\omega-
H_0+iQ}i\Phi\frac{1}{i\omega-H_0+iQ}i\Phi\right].
\label{ds2}
\ee
Clearly, $S_{\sigma}$ is nothing but the transport action for electrons in 
a random potential in the absence of the Coulomb interaction. 
The standard procedure \cite{finkelshtein} is to expand around the 
saddle-point of $S_\sigma$. The self-consistent saddle-point equation 
is given by:
\be
iQ(\br)=-g\langle\br|\frac{1}{i\omega-H_0+iQ} 
|\br\rangle.
\label{saddle}
\ee
The saddle-point solution is given by
\be
i[Q_0]_{\alpha\beta}^{mn}= q_0\delta_{\alpha\beta}\delta_{mn}+
{i\over 2\tau_0}\delta_{mn}\dl_{\al\beta}\mbox{sign}(n),
\label{q0}
\ee
where $\tau_0$ is the elastic scattering time. 
In general $\tau_0$ depends on the magnetic field $B$.
In the weak magnetic field limit, the Landau levels overlap due to disorder 
broadening and $\tau_0(B\to0)=1/2\pi \nu_0 g$ where $\nu_0$ is the
density of states.
In the strong field limit, $\omega_c\tau_0(0)\gg 1$, where $\omega_c$ is the 
cyclotron frequency. The diffusion comes from the ``skipping'' of the
semi-classical cyclotron orbits caused by impurity scattering.
One must solve Eq.~(\ref{saddle}) in the presence of quantized Landau
levels. Such a solution renders  
the famous semi-circle density of states by Ando\cite{ando}, 
usually referred to as the result of the SCBA.
Let us denote the saddle-point Green's function as
\be
[G_{sp}]_{\alpha\beta}^{mn}=\langle\v r\vert{1\over i\omega-H_0+iQ_0}
\vert\v r\rangle\delta_{\alpha\beta}\delta_{mn},
\label{gsp}
\ee
which is also called the SCBA Green's function.
Using $G_{sp}$ one can calculate the bare parameters of the theory, i.e.
the transport coefficients $\sigma_{xx}$ and $\sigma_{xy}$ at the level 
of SCBA. Without going into the details of the SCBA calculations, which can be
found in references \cite{ando,girvin,houghton,XRS}, we point 
out the following key results: 
1) in the strong field limit the elastic scattering 
time $\tau_0$ is of order $\sqrt{\tau_0(0)/\omega_c}$, the mean-free path 
$\ell=v_f \tau_0$ becomes the cyclotron radius $R_c$ and 
the diffusion constant 
$D={1\over2} R_c^2/\tau_0$ depends on the magnetic field; 
2) $\sigma_{xx}$ in the center of the $N$th Landau 
level is approximately given by
$\sigma_{xx}^{SCBA}\simeq(N+1/2)e^2/h$, 
therefore the perturbative expansion 
in $1/{\sigma}_{xx}$ is valid as long as $N>1$.  

The single-particle Green's function in Eq.~(\ref{saddle}), and hence the
saddle-point solution for $Q$ has, quite generally, a branch cut 
at $\omega=0$. Taking this into account, the saddle point solution assumes
the following generic form,
$iQ_0=q_0+ \frac{1}{2\tau}_0\Lambda$, where $\Lambda$ is a diagonal matrix,
$\Lambda_{\al\beta}^{nm}=\delta_{nm}\delta_{\al\beta } {\rm sign}
(n)$, in the space spanned by the replica ($\alpha,\beta$) and the 
Matsubara frequency ($n,m$) indices.
The non-linear $\sigma$ model is 
obtained by including the gapless, long-distance fluctuations around the saddle
point manifold of the form 
\be
{\tilde Q}(\v r)=U^{-1}(\br)Q_0 U(\br), 
\label{qtilde}
\ee
where $U$ is a unitary matrix $U\in U(M)$, and $M$ is the product of the 
number of replicas and that of the frequencies.
Defining a new dimensionless matrix field 
\be
Q(\v r)=U^{-1}(\v r)\Lambda U(\v r),
\label{q}
\ee
it is straightforward to show that
\bea
\Delta S_1&=&\pi \nu_0 Tr(Q\Phi)+i{1\over2}\Tr [(G_{sp}+G_{sp}^*)\Phi]+ 
{\cal O}(Q\nabla Q),
\label{s1q} \\
\Delta S_2&=&\pi \nu_0 \frac{\tau_0}{\hbar}\Tr(\Phi^2-Q\Phi Q\Phi) +\frac{1}{8}
\Tr[(G_{sp}+G_{sp}^*)\Phi (G_{sp}+G_{sp}^*)\Phi]+ {\cal O}(Q\nabla Q).
\label{s2q}
\eea
It is now convenient to denote the quantum average over $Q$ 
under the statistical weight $e^{-S_{\sigma}(Q)}$ by 
$\langle\ldots\rangle_{\sigma}$.
We obtain from Eqs.~(\ref{z2}-\ref{ds2}) the effective action,
\be
S_{\rm eff}(\Phi)=\langle\Delta S_1\rangle_{\sigma}+
\langle\Delta S_2\rangle_{\sigma}-
\frac{1}{2}\left[
\langle(\Delta S_1)^2\rangle_{\sigma}-\langle\Delta 
S_1\rangle_{\sigma}^2\right]+S_2^{(0)}(\Phi)+{\cal O}(\Phi^3)
\label{seff1}
\ee
The second and third $\Phi^2$ terms in Eq.~(\ref{seff1}) define the (density) 
polarization function $\Pi$,
\be
\langle\Delta S_2\rangle_{\sigma}-\frac{1}{2}\left[ \langle\Delta 
S^2_1\rangle_{\sigma}-
\langle\Delta S_1\rangle_{\sigma}^2\right]=\frac{1}{2}\sum_{n}\int 
d^2\br\int d\brp \Pi(\br,\brp;\omega_n)\Phi_{n}(\br)\Phi_{-
n}(\brp)
\label{pol}
\ee
The polarization function can be calculated in power series of 
$1/\sigma_{xx}$. 
To  leading order in $1/{\sigma}_{xx}$, we recover the result of the 
ladder-approximation:
\be
\Pi(\v q,i\omega_n)=\nu_0 \frac{Dq^2}{Dq^2+|\omega_n|}.
\label{Pi}
\ee
Higher order interference corrections presumably renormalizes the diffusion 
constant $D$, and the thermodynamic DOS $\nu_0\to dn/d\mu$ in Eq.~(\ref{Pi}).
Thus, we have derived the effective action to order $\Phi^2$, 
\be
S_{\rm eff}\approx\frac{T}{2}\sum_{n}\Phi_n(\br)\left[v^{-1}(\br-\brp)+\Pi(\br-
\brp;i\omega_n)\right]\Phi_{n}(\brp).
\label{eff-action}
\ee
The scalar field $\Phi$ precisely mediates
the diffusion-screened electron-electron Coulomb
interaction. \cite{aal}

\subsection{Semi-classical phase-approximation}

Now we turn to the evaluation of the impurity averaged single
particle Green's function given in Eq.~(\ref{avgG2}).
By a simple reordering of functional integrals,\cite{KimWen,qcgap}
we have
\be
\langle G(\v r,\v r,\tau)\rangle =\int D[\Phi]
\ov{G}(\v r,\v r,\tau,\Phi)e^{-S_{\rm eff}(\Phi)},
\label{avgG3}
\ee
where $\ov{G}$ is the averaged Green's function in a fixed configuration
of the scalar potential $\Phi$,
\be
\ov{G}(\v r,\v r, \Phi)=\lim_{n\rightarrow 0}\int D[Q]\int D  \ov{\psi} D\psi 
\psi_{\al_1}\psi^*_{\al_1 } 
e^{-S({\psi^*},\psi, Q,\Phi)}/{\cal Z}[\Phi].
\label{gphi}
\ee
We now make an approximation regarding ${\ov G}(\Phi)$ that takes into
account exclusively the important interference effects between the
phases of the electron wave functions. The amplitude fluctuations
are small for the slowly varying fluctuations of the $\Phi$-field
that dominate the contributions to the effective action in 
Eq.~(\ref{eff-action}). Since these fluctuations are spatially
smooth on the scale of the elastic mean-free path $\ell$, i.e.
$\nabla \Phi \ell/E_f \ll 1$, they do not significantly alter the 
classical trajectory of the electrons. This is a unique feature
of the slow diffusive dynamics of the electrons in a random media.
Appealing now to the semi-classical 
approximation, the single-electron propagator in the presence of
interactions is modified by a U(1) phase factor:
\be
{\ov G}(\br,\br;\tau;\Phi)\approx {\ov G}(\br,\br;\tau;0) e^{-\Delta 
S_{cl}(\Phi)}.
\label{gphicl}
\ee 
where $\Delta S_{cl}$ is the change of the action caused by sampling
the potential $\Phi$ along the classical path:
\be
\Delta S_{cl}(\Phi)=-i\int_0^\tau d\tau^\prime
\Phi[\br^\prime _{cl}(\tau^\prime ),\tau^\prime],
\label{dscl}
\ee
with $\br^\prime _{cl}(\tau^\prime )$ being the classical 
trajectory that starts and ends at $\br$ in the presence of random 
potential but in the  absence of the $\Phi$-field. 
Upon averaging over the random potentials, the classical trajectory 
can be described by a random walk. Let $\rho(\br,\tau)$ be the 
probability of a particle being at $\br$ at time $\tau$,
\be
\Phi(\br^\prime _{cl}(\tau^\prime ),\tau^\prime ) 
=\int d\br^\prime  \rho(\br^\prime ,\tau^\prime ) 
\Phi (\br^\prime ,\tau^\prime ).
\label{phicl}
\ee
Since the critical conductivity is finite at the IQHT, the charge
spreading is expected to be described by (anomalous) diffusion.
The probability density 
$\rho$ then satisfies the following diffusion equation,
\be
[-D(\nabla^\prime )^2 +\partial_{\tau^\prime }]\rho(\br^\prime ,\tau^\prime )=[\dl(\tau^\prime )-\dl(\tau^\prime -
\tau)]\dl(\br^\prime -\br),
\label{diffusion}
\ee
where the $\delta$-functions on the right hand side result from the
boundary conditions imposed on the original trajectory and correspond
to injecting an electron at $\v r$ and time $0$ and removing it at
time $\tau$. The associated current density is
given by
\be
{\v J}=-D(\nabla-\gamma_H {\hat z}\times\nabla)\rho,
\label{diffusioncurrent}
\ee
where $\gamma_H=\sigma_{xy}/\sigma_{xx}$ is the Hall ratio.
Notice that $\gamma_H$ does not enter the diffusion equation
(\ref{diffusion}) because the transverse force does not
affect the charge spreading  which
is described by $\nabla\cdot\v J$ in the continuity equation.
Solving Eq.~(\ref{diffusion}) {\it in the bulk} of systems
without edges, we find,
\be
\rho(q,i\omega_n)=\frac{1-e^{i\omega_n\tau}}{Dq^2+|\omega_n|}.
\label{rho}
\ee
Later we will show that $\gamma_H$ does enter in
the presence physical edges. In that case, the diffusion
equation must be solved with the appropriate spatial boundary conditions.

Inserting the results of Eqs.~(\ref{rho}), (\ref{dscl}) and (\ref{phicl}) into
Eq.~(\ref{gphicl}), we have
\be
{\ov G}(\br,\br;\tau;\Phi) \approx  \ov{ G}(\br,\br;\tau;0)  
e^{{ i\int_0^\tau d\tau^\prime \int d\br^\prime  
\Phi(\br^\prime,\tau^\prime )\rho(\br^\prime,\tau^\prime)}}.
\label{phase-approx}
\ee
Note that the above is but a special case of the more general phase 
approximation in the presence of U(1) 
gauge-fields.\cite{KimWen,shytov,qcgap}  
The quantum interference
effects can be included by the renormalization of the diffusion
constant $D$ and other parameters of the theory.
In fact it has been argued recently in reference\cite{alex} that 
the phase approximation of Eq.~(\ref{phase-approx}), along with the effective 
action of the 
screened-potential of Eq.~(\ref{eff-action}) can be derived by seeking a 
temporally and spatially varying saddle-point solution $Q_0(\br,\tau,\Phi)$  
of the action $S_{Q}(\Phi)$ for each $\Phi(\br,\tau)$. Quantum interference 
can be treated systematically by considering fluctuations around such saddle 
point solutions. We note in passing that, at criticality,
another complication arises: 
due to the multifractal behavior
of the critical eigenstates, the diffusion 
is anomalous, i.e., $D$ becomes dependent on the length and the time scales.  
These subtleties will be addressed in later sections.

The final step is to substitute Eq.~(\ref{phase-approx})
into Eq.~(\ref{avgG3}) and carry out the functional integral.
Taking into account the Gaussian fluctuations
in $\Phi$ captured by the effective action in Eq.~(\ref{eff-action}),
we obtain the central result: 
\be
\langle G(\br,\br;\tau)\rangle \approx \ov{G}(\br,\br;\tau;0) 
e^{-W(\tau)}
\label{finalG}
\ee
where the Debye-Waller phase-delay factor is
\be
W(\tau)=\frac{T}{2}\sum_{n}\int{d^2q\over(2\pi)^2}\rho(\v q,i\omega_n)
v_{\rm sc}(q,i\omega_n)\rho(-\v q,-i\omega_n),
\label{finaldw}
\ee 
and $v_{\rm sc}$ is the dynamically 
screened interaction implied in the effective action 
in Eq.~(\ref{eff-action}),
\be
v_{\rm sc}(\v q,i\omega_n)=\frac{v(q)}
{1+v(q)\Pi(q,\omega_n)}=
\frac{v(q)}{1+{dn\over d\mu} v(q)\frac{Dq^2}{Dq^2+|\omega_n|}}.
\label{vsc}
\ee
In Eq.~(\ref{finalG}), $\ov{G}(\tau)$ corresponds to the SCBA Green's function 
$G_{sp}(\tau)$ defined in Eq.~(\ref{gsp}),
\be
\ov{G}(\v r, \v r, \tau)=-i\pi\nu_0{1\over\beta}\sum_n
e^{-i\omega_n\tau}{\rm sign}(\omega_n)
=-\nu_0{\pi/\beta\over \sin(\tau\pi/\beta)}.
\label{scbag}
\ee

After carrying out the sum over Matsubara frequency in 
Eq.~(\ref{finaldw}), the details of which are given in Appendix A, we obtain
the finite-temperature expression:
\be
W(\tau)=\int_{-\infty}^{+\infty}{d\omega\over2\pi i}[f(-i\omega)-f(i\omega)]
\frac{1-e^{\omega\tau}}{e^{\beta\omega}-1},
\label{finaldw2}
\ee
where 
\be
f(-i\omega)=\frac{1}{2}\int {d^2 q\over(2\pi)^2}
\left(\frac{1}{Dq^2-i\omega}\right)^2 v_{\rm sc}(q,-i\omega).
\label{f(omega)}
\ee
The interaction correction to the TDOS is determined by the
behavior of the phase-factor $W(\tau)$, which depends on the
nature of the dynamically screened interaction $v_{\rm sc}$. 

\section{Bulk tunneling density of states in 2D}

We now derive the TDOS at $T=0$.
It is necessary to perform the following 
analytical continuation:
\be
\nu(\omega)=-\frac{1}{\pi} {\rm Im} 
G(\br,\br; i\omega_n)\vert_{i\omega_n\rightarrow
\omega+i\dl}=-{1\over\pi}{\rm Im}\left[\int d\tau
e^{i\omega_n\tau} \ov{G}(\v r,\v r, \tau)e^{-W(\tau)}
\right]_{i\omega_n\rightarrow\omega+i\dl}.
\label{analytic}
\ee
The procedure turns out to be quite non-trivial. 
Since we could not find discussions of the 
technique in the literature, we elect to include the details of the
analytical continuation in Appendix B, where we show, in the limit
$T\to0$,
\be 
\nu(\omega)\approx \frac{2}{\pi}\nu_0
\int_0^{\infty}\frac{\sin(|\omega| t)}{t}e^{-
W(it)},
\label{nuomega}
\ee
where $\nu_0$ is the noninteracting TDOS near the Fermi level.
Taking the $T=0$ limit of Eq.~(\ref{finaldw2}), we get
\be
W(it)=\int_0^{\infty} {d\omega\over2\pi i}
[f(-i\omega)-f(i\omega)]
(1-e^{-i\omega t}),
\ee
with the function $f$ given in Eq.~ (\ref{f(omega)}). 
The term with the oscillatory factor
$e^{-i\omega t}$ averages to zero upon integration
except for $\omega\ll 1/t$ where 
$e^{-i\omega t}\approx 1$. Therefore we can effectively leave out the $e^{-
i\omega t}$ term and introduce a lower cut-off $\hbar/t$ to the integral: 
\be
W(it)=\int_{1/t}^{1/\tau_0} d\omega \frac{1}{2\pi i}[f(-
i\omega)-f(i\omega)].
\ee
The upper cut-off of the integral in the above equation
arises from the fact that the diffusive picture 
becomes invalid at time scales shorter than the elastic scattering time.
Using Eq.~(\ref{f(omega)}), we obtain,
\be
W(it)=\int_{1/t}^{1/\tau_0} {d\omega\over2\pi}
\int{d^2q\over(2\pi)^2}{\rm Im}\left[ v_{\rm sc}(q,-i\omega)
({1\over Dq^2-i\omega})^2\right].
\label{dwintegral}
\ee
We next turn to the evaluation of the most singular contributions
to $W$ and thus to the TDOS for different forms of interactions.

\subsection{Long range Coulomb interaction}

The singularity in the TDOS  arises from the physics of 
dynamical screening. For Coulomb interaction,
$v(q)=2\pi e^2/q$. The dynamical screened interaction in Eq.~(\ref{vsc})
becomes,
\be
v_{\rm sc}(q,-i\omega)={2\pi e^2\over q+{\kappa D q^2\over Dq^2-i\omega}},
\label{vsccoulomb}
\ee
where $\kappa=2\pi e^2 dn/d\mu$ is the inverse screening length at the
transition. It is important to note that in the presence of disorder,
the range of validity for static screening is quite small.\cite{palee}
Since diffusion is a relatively slow process, at non-zero 
frequency, the long distance singularity associated with
the long range Coulomb interaction is not screened, as
can be seen from Eq.~(\ref{vsccoulomb}). In fact, in the region
where $Dq^2<\omega < D\k q$, the effective interaction has the most
singular form:
\be
v_{\rm sc}(q,-i\omega)\sim\frac{1}{D\k q^2},\quad
Dq^2<\omega < D\k q,
\label{vscsingular}
\ee
which gives the main contribution to the wave-vector integral in
Eq.~(\ref{dwintegral}),
\be
W(it)={1\over4\pi^2\sxx}
\int_{1/t}^{1/\tau_0}{d\omega\over\omega}\ln({\omega\over D\kappa^2}).
\label{dwomegaintegral}
\ee
Note that in this region, the diffusion coefficient $D$ is a constant. The
anomalous diffusive behavior \cite{cd} in the regime $Dq^2\gg\vert\omega\vert$
does not affect the leading contribution.
The remaining frequency integral generates the 
double-logarithmic dependence in time:
\be
W(it)\approx \frac{1}{8\pi^2\sxx}\ln  
({t\over\tau_0})\ln ({t\over\tau_1}),
\label{dwcoul}
\ee
where $\tau_1=1/\tau_0(D\kappa^2)^2$, and
$\sxx=D dn/d\mu$ is the conductivity defined via
the Einstein relation. Near the Landau
level centers, one can show in the SCBA that
$\tau_1/\tau_0=(1/4\pi^4\sxx) (k_f a_B)^2/k_f\ell\ll1$.
This double-logarithmic form
is the dominant behavior of the Debye-Waller
phase factor in the long-time limit. Next order corrections
are of the order $\{1/\sigma_{xx},1/\sigma_{xx}^2\}\ln (t/\tau_0)$.
The contributions from all 
six different integration regions in the $(\omega,q)$-plane
are discussed in detail in Appendix C.
Substituting Eq.~({\ref{dwcoul}) into Eq.~(\ref{nuomega}), we obtain
the zero-temperature TDOS in the Coulomb case,
\be
\nu(\omega)={2\nu_0\over\pi}\int_0^\infty dt
{\sin(\vov t)\over t}
e^{{-{1\over8\pi^2\sxx}\ln ({t\over\tau_0})\ln ({t\over\tau_1})}}.
\label{nuomegacoul}
\ee
Keeping in mind that in deriving this result we have assumed
a frequency independent conductivity $\sigma_{xx}$, i.e. we have
neglected the quantum interference effects. In general, $\sxx$
is renormalized by localization effects,
of leading order $(1/\sxx)\ln\omega\tau_0$ in the unitary ensemble,
and by interaction effects of leading order $\ln\omega\tau_0$
in strong magnetic field.\cite{girvin,houghton} Thus $\sxx$ takes on
the frequency-independent SCBA value only if
$\vert\ln(\ve\tau_0)\vert\ll\sxx$.

\subsubsection{High frequency regime: $\vert\ln(\omega\tau_0)\vert\ll\sxx$}

In this regime, the weak localization correction to the conductivity
can be neglected. If in addition, $\vert\ln(\omega\tau_0)\vert\ll\sqrt{\sxx}$,
we can expand the exponential in Eq.~(\ref{nuomegacoul}) to leading
order in $1/\sxx$ and obtain,
\be
\nu(\omega)=\nu_0\left[1-{1\over 8\pi^2\sxx}\ln(\vov\tau_0)\ln(\vov\tau_1)
\right].
\label{nupert}
\ee
This reproduces the high-field perturbative diagrammatic result of
Girvin, Jonson, and Lee \cite{girvin} and Houghton, Senna, and Ying.
\cite{houghton} For frequencies in the range,
$\sqrt{\sxx} \ll \vert\ln(\vov\tau_0)\vert \ll \sxx$,
the integral in Eq.~(\ref{nuomegacoul}) can be evaluated by the 
stationary point/instanton method, leading to a nonperturbative
resummation of the double-log divergences in Eq.~(\ref{nupert}),
\be
\nu(\omega)=\nu_0\exp\left[-{1\over8\pi^2\sxx}\ln(\vov\tau_0)
\ln(\vov\tau_1)\right].
\label{nuhighf}
\ee
In the zero magnetic field case, such a
nonperturbative resummation of the perturbative double-log divergences 
was carried out by Finkel'stein, \cite{finkelshtein} and recently
reexamined using different approaches. \cite{levitov,alex,kop}
Our result of Eq.~(\ref{nuhighf}) can be regarded as an extension
of the latter to the strong magnetic field case.

\subsubsection{Low frequency regime: $\vert\ln(\ve\tau_0)\vert\gg\sxx$}

Here the quantum interference effects will, in general, lead to a frequency
dependent conductivity. However, at the IQHT,
the critical conductivity $\sxxc$ is finite and of the order
of $e^2/\hbar$. This experimental fact has been shown
numerically for both noninteracting electrons and interacting electrons
in the HF theory.\cite{aoki,cd,huo,wjl,yang2,thf}
Thus, the validity of our analysis, i.e. the structure of the double-log
divergence at long-times, can be extended into the regime of small
$\omega$, provided that $\sxx$ in Eq.~(\ref{nuomegacoul}) is replaced
by the critical conductivity $\sxxc\simeq0.5/2\pi$.
Note that the due to the double-log term in the exponent,
Eq.~(\ref{nuomegacoul}) implies
$$\lim_{t\rightarrow \infty}W(it)= +\infty,$$ 
and consequently a zero-bias anomaly in the TDOS
$$\nu(\omega=0)=0.$$

To obtain the limiting behavior of $\nu(\omega)$ for small $\omega$,
we expand the $\sin(\omega t)$ factor in Eq.~(\ref{nuomegacoul})
in a power series in $\omega t$.
It is important to emphasize that
this is possible because of the
double-log contribution which makes the time integral over $e^{-
W(it)}$ converge fast enough such that the TDOS becomes analytic at
small $\omega$. The claims made by Polyakov and Samokhin
\cite{polyakov} that the TDOS falls off faster 
than any power law in $\omega$ is in fact incorrect.
Since the signs of the expansion-coefficients alternate, the series is 
asymptotic, i.e., it can be infinitely accurate at small $\omega$.  
To first order in $\omega$,  
\be
\nu(\omega)=\nu_0\vov{2\over\pi}\int_0^\infty dt 
e^{{-{1\over8\pi^2\sxxc}\ln ({t\over\tau_0})\ln ({t\over\tau_1})}}.
\ee
Performing this integral, and use the fact that the compressibility is
only weakly renormalized, {\it i.e.} $dn/d\mu\simeq \nu_0$, 
we obtain the 2D quantum Coulomb gap behavior given
in Eq.~(\ref{Coulomb}) in the introduction, i.e.
\be
\nu(\omega)=C_Q \hbar\vert\omega\vert/e^4.
\label{Coulomb2}
\ee
In contrast to the 2D classical Coulomb gap, the slope $C_Q$ is
not a universal number, but rather a quantity of quantum mechanical
origin that depends on the microscopic details of the sample,
\be
C_Q=\sqrt{\frac{1}{2\pi^3\sxxc}}\left[1+{\bf \Phi}(\sqrt{
2\pi^2\sxxc}) \right] 
e^{(2\pi^2\sxxc)+(1/8\pi^2\sxxc)\log ^2R},
\label{cq}
\ee
where ${\bf \Phi(x)}$ is the error 
function and $R=\sqrt{\tau_1/\tau_0}=1/D\kappa^2\tau_0$ is
a quantity that depends on the degree of disorder. The latter can
be written in terms of more familiar quantities according to
\be
R={1\over4\pi^4\sxxc}{(k_f a_B)^2\over k_f\ell_0},
\ee
where $a_B$ is the Bohr radius and $\ell_0$ is the {\it zero-field} 
mean free path.
\begin{figure}    
\center    
\centerline{\epsfysize=2.8in    
\epsfbox{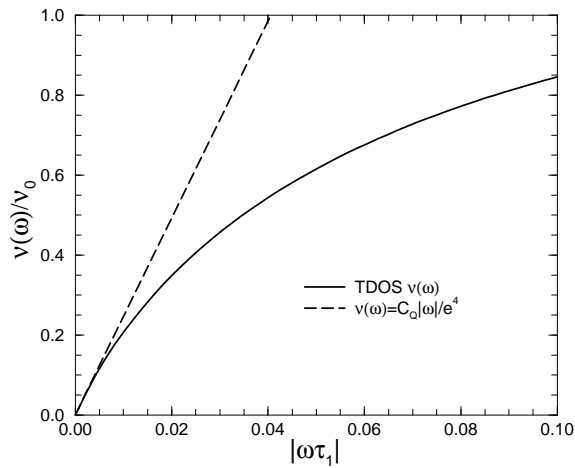}}
\begin{minipage}[t]{8.1cm}     
\caption{The TDOS in Eq.~(\ref{nuomegacoul})
in the case of long-range Coulomb interaction, showing
the asymptotic linear Coulomb pseudogap behavior at small $\vov$.
The parameters are $\sxx=\sxxc=0.5e^2/h$ and $\tau_0/\tau_1=10$.
}   
\label{fig2}      
\end{minipage}     
\end{figure}    
It is easy to verify that the next term in the expansion
is of the order $\omega (\omega\tau_0)^2e^{\sigma_{xx}}$,
which is small in this regime. The results of numerical 
integration of Eq.~(\ref{nuomegacoul})
is plotted Fig.~2, which shows the crossover from the
high frequency behavior described by Eq.~(\ref{nuhighf}) to
the asymptotic linear Coulomb gap of Eq.~(\ref{Coulomb2}) at low frequencies.
Since the real transition must be governed by an interacting
fixed point where the Coulomb interaction strength is finite,
we conclude that the true asymptotic behavior of the bulk TDOS exhibits
the quantum Coulomb gap at the IQHT.

\subsection{Short-range interactions}

In this subsection, we address the question of how short range interactions,
such as screened Coulomb interactions, which are irrelevant perturbations
at the NIFP in the RG sense, cause depletion of the TDOS near the Fermi level.
For simplicity, we focus on the local interactions described by the
prototype short range interacting potential $v(\br-\brp)=u\dl(\br-\brp)$ 
and $v(q)=u$. The screened interaction in Eq.~(\ref{vsc}) becomes,
\be
v_{\rm sc}(q,i\omega_n)={u\over 1+ u{dn\over d\mu}{Dq^2\over Dq^2+\vnv}}.
\ee
Inserting this expression into the Eq.~(\ref{dwintegral}) for $W(it)$,
one finds,
\be
W_{\rm sr}(it)=\int_0^{1/\tau_0}{d\omega\over2\pi}
(1-e^{-i\omega t})
\int {d^2q\over(2\pi)^2}
{\rm Im}\left[{1\over Dq^2-i\omega}{1\over D^\prime q^2-i\omega}
\right],
\label{dwintegralsr}
\ee
where $D^\prime=D+u\sxx$.
In contrast to the long range Coulomb case, the contributions 
to the $q$-integral from the $\omega>Dq^2$ regime and the 
$\omega<Dq^2$ regime are now comparable. 

\subsubsection{Pre-scaling regime}

Let us first ignore the quantum interference effect and focus on
the perturbative regime appropriate when $\vert\ln\omega\tau_0\vert
\ll\sqrt{\sxx}$. 
In this case we can treat the diffusion coefficient $D$ as a 
constant, and the interaction strength $u$ as a marginal perturbation
(a scale invariant constant) in Eq.~(\ref{dwintegralsr}). 
Carrying out integrations, we arrive at,
\be
W_{\rm sr}(it)=\int_0^{1/\tau_0}d\omega{\alpha\over\omega}(1-e^{-i\omega t})
\label{dwsr}
\ee
where $\alpha$ is a non-universal dimensionless quantity dependent
on the interaction strength,
\be
\alpha=\lambda{1\over8\pi^2\sxx}{2+\lambda\over (1+\lambda)^2}
\left(1+\ln\sqrt{1+\lambda}\right),
\label{alpha}
\ee
with $\lambda=u\nu_0$. Thus 
$W(it)$ diverges in the long time limit only logarithmic ally.
The situation is completely analogous to the classic X-ray edge problem.
\cite{mahan} Taking care of the short-time behavior in Eq.~(\ref{dwsr})
as in the X-ray edge problem, we obtain,
\be
W(it)\simeq=\alpha\ln(1+t/\tau_0).
\label{W-short}
\ee
Substituting Eq.~(\ref{W-short})
into Eq.~(\ref{nuomega}), we have for the TDOS
\be
{\nu_{\rm sr}(\omega)\over\nu_0}={2\over\pi}\nu_0\int_0^{\tau_0}dt 
{\sin(\vert \omega t)\over t}{1\over (1+t/\tau_0)^\alpha}
=C_\alpha \vert\omega\tau_0\vert^\alpha,
\label{nusrprescaling}
\ee
where $C_\alpha=(2\pi)\int_0^\infty dy {\sin y\over y}{1\over (1+y)^\alpha}$
is a dimensionless numerical constant. Thus we conclude that a nonuniversal
power-law suppression of the TDOS prevails in the prescaling regime for
short-range interactions.

\subsubsection{Scaling regime}

On approaching the scaling regime of the IQHT,
it is necessary to take into account the scaling behaviors of
(1) the diffusion coefficient $D$ and (2) the interaction strength $u$.
It is known from the work of Chalker and Daniel \cite{cd}
that the multifractality of the critical eigenstates 
leads to anomalous diffusion in the regime $Dq^2> 
\omega$. The diffusion constant becomes a function of $q^2/\omega$:
 \be
D(q,\omega)= D (qL_{\omega})^{-\eta}, 
\ee
where as before, $L_\omega=\sqrt{D/\omega}$ and $\eta=-x_-$
which is given in Eq.~(\ref{x-}), and
$D_2=2-\eta$ is the multifractal dimension.
The critical conductivity $\sxxc$ is once again finite and will be
taken as scale-independent.
The Debye-Waller phase factor $W_{\rm sr}(it)$ 
in Eq.~(\ref{dwsr}) is now modified by replacing $\alpha\to\alpha^\prime$
where
\be
\alpha^\prime={1\over8\pi^2\sxxc}\left[c_\eta+\log(1+\lambda)
\right],
\label{alphaprime}
\ee
with $c_\eta=1/2+2/(4-3\eta)$.
This modification due to the multifractal behavior alone would lead to, 
repeating the calculations above, the same behavior of the TDOS
as in Eq.~(\ref{nusrprescaling}) in the prescaling regime, 
except the exponent $\alpha$ is replaced by $\alpha^\prime$.

Next we must take into account the fact that $u$ is an irrelevant
perturbation. The effective interaction scales to zero according
to \cite{lwinter,wangetal} $u_{\rm eff}\sim u\omega^{x_+/z}$, where
$-x_+$ is the scaling dimension for short-range interactions
discussed in section II and $z=2$ is the dynamical exponent at the
NIFP. As a result, the quantity $\alpha^\prime$ obeys the following
scaling relation,
\be
\alpha^\prime(u,\omega)={\cal A}(u\omega^{x_+/z}).
\label{alphascaling}
\ee
The fact that
${\cal A}(u\to0,\omega)=0$ implies, together with Eq.~(\ref{alphaprime}),
the leading scaling behavior for $\alpha^\prime$,
\be
\alpha^\prime\simeq A \lambda (\omega\tau_0)^{x_+/z},\quad
A=c_\eta /4\pi^2\sxxc.
\label{leadingalpha}
\ee
Substituting this result into Eq.~(\ref{dwsr}), we find
\be
W_{\rm sr}(it)=A \lambda \gamma^{-1}\left[\left({\tau_0\over t}
\right)^{\gamma}-1\right],
\label{dwscaling}
\ee
where $\gamma=x_+/z\simeq0.32$. 
That $W_{\rm sr}(it)$ converges now in the
long-time limit should be contrasted to the long range Coulomb case and
is a consequence of the short range interactions being irrelevant,
i.e. $\gamma>0$. An immediate implication is that the TDOS would
be finite at zero bias and the level spacing scales as $1/L^2$
as in the noninteracting theory. However, we shall show below that
if the bare interaction $\lambda$ is large, it still leads to strong
suppression of the TDOS at low energies.

From Eqs.~(\ref{nuomega}) and (\ref{dwscaling}), the TDOS is given by
\be
\nu_{\rm sr}(\omega)=\nu(0){2\over\pi}\int_{\tau_0}^\infty dt
{\sin(\vov t)\over t} e^{(A\lambda/\gamma)(\tau_0/t)^\gamma},
\label{nusrscaling}
\ee
where $\nu(0)=\nu_0 e^{-A\lambda/\gamma} < \nu_0$.
Performing the integral using the saddle-point/instanton approximation,
we find that at low frequencies, $\omega < \omega_0$,
the TDOS is given by
\be
\nu_{\rm sr}(\omega)=\nu(0)\left[1+\left({\vov\over\omega_0}
\right)^\gamma\right],
\label{nusrfinal}
\ee
where $\omega_0=\tau_0^{-1}(A\lambda/\gamma)^{-1/\gamma}$ is an energy 
scale. We see that upon scaling, the irrelevance of short-range interactions
leads to a smearing of the power-law behavior in the perturbative regime,
giving rise to a finite and nonuniversal TDOS at zero bias. However,
although short-range interactions are irrelevant in the RG sense,
since $\nu(0)\ll\nu_0$ for large $\lambda$, they still lead
to strong density of states suppression at low bias. 
What is remarkable is that Eq.~(\ref{nusrfinal}) predicts an increase
of the TDOS with energy that follows a {\it universal} power law,
with an initial cusp singularity for our value of $\gamma$.
These predictions can, in principle, be tested experimentally by
deliberately screening out the long-ranged Coulomb interaction
using metallic gates or ground planes.

\subsection{General interacting potential: $ v(q)=\frac{u}{q^{2-p}}$}

It is interesting to
consider a general interacting potential of the form 
$v(q)=\frac{u}{q^{2-p}}$. We find that for $p<2$, the dominant 
contribution to $W(it)$ in the long time limit comes from the same double-log
term as in the case of Coulomb interaction (corresponding to $p=1$). We 
conclude that for $p<2$, the density of states in the asymptotic
$\omega\rightarrow 0$ limit is of the form of the linear gap
$\nu(\omega)\sim \omega$. For $p>2$, the phase-delay factor 
$W(it)$ approaches a constant $W_{\infty}>0$ for $t \gg\tau_0$. 
In this case the density of states does not vanish, but rather 
develops a shallow dip at 
$\omega=0$, where $\nu(0)=\nu_0 e^{-W_{\infty}}$.
The borderline case is that of $p=2$, corresponding to the   
$\delta$-function interaction studied. It can be shown from either the $p>2$ 
or the $p<2$ side that as $p\rightarrow 2$, a single-log term 
emerges and dominates the contributions in $W(it)$, leading to the power 
law density of states of Eq.~(\ref{nusrprescaling})
in the perturbative regime, and to the finite zero bias TDOS
we obtained in Eq.~(\ref{nusrfinal}). Details of this analysis can
be found in Appendix C.

\section{TDOS in quasi-1D systems with edges}

In the cases studied above, the {\it bulk} TDOS
does not depend on the Hall conductance. This is in keeping 
with the fact that the bulk diffusion equation
is the same with or without time reversal 
symmetry. The traverse force induced by a magnetic field
does not affect the diffusive charge spreading.
It is well-known that in the non-interacting theory
of the IQHT, the term in the action that depends on $\sigma_{xy}$ is 
topological and non-perturbative.\cite{pruiskenbook}
However, it has been discovered
recently that in the presence of edges, the Hall 
conductance enters measurable quantities even in the perturbative limit. The 
topological term gives rise to a tilted  boundary condition 
for diffusion and at more subtle levels affects the quantum interference 
processes. For 
example it was shown by Khmel'nitskii and Yosefin,\cite{KY} and by
Xiong, Read and Stone\cite{XRS} that the mesoscopic conductance fluctuations 
in phase-coherent samples become dependent on the Hall conductance in 
the presence of edges. More recently 
Shytov, Levitov, and Halperin demonstrated that 
the I-V curves for {\it edge tunneling} into the 1D Luttinger liquid like 
edge excitations of fractional quantum Hall liquids can be obtained from
the point of view of bulk composite fermions by using a similar phase 
approximation in treating the effects of gauge fluctuations.\cite{shytov}
There, the $\sigma_{xy}$ dependence in the exponent of the power-law tunneling
conductance also arises from  a boundary condition of the 
source current at the edge of a semi-infinite sample. 
Our case differs from and is simpler than that of 
the composite fermions in the sense that we need to 
consider only pure potential fluctuations mediated interactions
in the integer quantum Hall regime. 

To study how the physical boundaries bring the Hall ratio
into the {\it bulk} TDOS, we consider, instead of the half plane geometry,
\cite{shytov} a quasi-1D sample with its length $L$ much 
greater than its width $W$, exposing two reflecting edges along its width. 
This condition can be realized experimentally in the long Hall bar geometry,
and is the same as that considered by Xiong, Read and Stone \cite{XRS} in
their study of the edge effects on mesoscopic conductance fluctuations
in strong magnetic fields. 

Because the incident current is at an angle with the reflecting edges,
the presence of the magnetic field affects the diffusion 
process through a modified boundary condition which 
depends on the Hall ratio $\gamma_H=\sxy/\sxx $,
\be
\left[\partial_n 
+\gamma_H\partial_t\right]\rho=0.
\label{bc}
\ee
Here the subscript $n$ denotes the directions normal to the edge and 
$t$ denotes the tangential direction.
Strictly speaking, if the new boundary conditions (\ref{bc})
are taken into account, the diffusion propagator as well as the screened 
interactions will depend on the Hall conductivity. Such effect is minimal 
if the sample is wider than it is long and becomes pronounced only 
in the quasi-1D limit when $L\gg W$. 
For simplicity, we consider the case of the $\delta$-function 
interaction encountered in the previous section and the geometry of an 
infinite strip with hard walls at $y=0$ and $y=W$. We also limit ourselves
to the perturbative regime, and neglect scaling corrections to the
conductivities and the interaction strength.

The diffusive modes that are solutions of the diffusion equation
(\ref{diffusion}) and satisfy the boundary condition (\ref{bc}) 
can be obtained as follows,\cite{XRS}
\bea
\phi_{k,q}^{L}&=&a_k e^{iqx}\left[\frac{k\pi}{W}\cos(\frac{k\pi y}{W})-
i\gamma_H q \sin(\frac{k\pi y}{W})\right],\quad\mbox{for } k\neq 0\non\\
\phi_{0,q}^{L}&=&a_0 e^{iqx-i\gamma_H q y}, \quad\mbox{for } k= 0.
\label{lmodes}
\eea
Since the boundary condition is not self-adjoint, there are also a set 
of right eigen-functions (with the same eigen-values)
that satisfy the boundary condition under parity transformation:
\be
\left[\partial_n 
-\gamma_H\partial_t\right]\phi^{R}=0. \label{bcr}
\ee
They are given by,
\bea
\phi_{k,q}^{R}&=&a_k e^{iqx}\left[\frac{k\pi}{W}\cos(\frac{k\pi 
y}{W})+i\gamma_H q \sin(\frac{k\pi y}{W})\right],\quad\mbox{for } k\neq 
0\non\\
\phi_{0,q}^{R}&=&a_0 e^{iqx+i\gamma_H q y}, \quad\mbox{for } k= 0.
\label{rmodes}
\eea
The $\{\phi^L,\phi^R\}$  are the eigenmodes of the 
Laplacian operator, 
\be
-D\nabla^2 \phi^{L,R}_q(x,y)=\Lambda_{k,q}\phi^{L,R}_q(x,y);
\ee
where 
\bea
\Lambda_{q,k}&=&D\left[\frac{k^2\pi^2}{W^2}+q^2\right],\quad\mbox{for } 
k\neq 0\non\\
\Lambda_{q,0}&=&D(1+\gamma_H^2)q^2,\quad \mbox{for } k= 0.
\label{modes}
\eea
Using the bi-orthogonality relation and the completeness condition,
we can express the Debye-Waller phase factor $W(it)$ in Eq.~(\ref{dwintegral})
in terms of the sum over the eigenmodes in the transverse channels:
\be
W(it)=\frac{1}{2}\int_{1/t}^{\tau_0} {d\omega\over(2\pi)} \sum_{q,k} 
\phi^L_{k,q}(\br)[\phi^R_{k,q}(\br)]^* 2{\rm Re}\left\{
\frac{1}{(\Lambda_{k,q}-i\omega)^2}\frac{u}{1+u\nu_0\frac{\Lambda_{k,q}
}{\Lambda_{k,q}-i\omega}}\right\}.
\ee
In the limit $Dt\gg W^2$, contributions 
from $k>0$ modes can be ignored. For the consideration of $\nu(\omega)$, this
condition translates into $L_{\omega}\gg W$, i.e. $\hbar\omega>>D/W^2$.
Strictly speaking, $W(it)$ becomes dependent on 
the spatial position, but we do not expect any spatial singularity. 
It is therefore justifiable to average $W(it)$ over the entire strip. 
We obtain,
\be
W(it)\approx 
\frac{4}{3\pi^2}\frac{u}{\hbar\sqrt{D_1}}{1\over 1+\lambda}(\sqrt{t}-
\sqrt{\tau_0}),
\label{dw1d}
\ee
where $D_1=(1+\gamma_H^2) D$.
Thus, the phase factor is dominated in quasi-1D by the 
$\sqrt{t}$ divergence in the long-time limit.
Rewrite the prefactor in Eq.~({\ref{dw1d}) as
\be
\sqrt{\omega_{B}}=\frac{4}{3\pi^2}\frac{e\sqrt{\nu_0}}{\hbar^{3/2}
\sqrt{(1+\gamma_H^2)\sxx}}\frac{\lambda}{1+\lambda},
\label{omega01d}
\ee
we have $W(it)=\sqrt{\omega_B t}-\sqrt{\omega_B\tau_0}$.
Upon substitution of $W(it)$ into 
Eq.~(\ref{nuomega}), the TDOS in the quasi-1D case is given by,
\be
\nu_{\rm q1D}(\omega)=\nu_0{2\over\pi}\int_{\tau_0}^\infty dt
{\sin(\vov t)\over t}e^{-\sqrt{\omega_B}(\sqrt{t}-\sqrt{\tau_0})}.
\label{nu1d}
\ee
Approximating the time-integral by the stationary point/instanton
solution, we find that for $\omega\ll\omega_B$,
\be
\nu_{\rm q1D}(\omega)\propto \nu_0e^{-{\omega_B\over4\vov}}.
\label{nu1dfinal-1}
\ee
Thus, the TDOS is strongly suppressed at low energies in quasi-1D
systems in a manner that is sensitive to the applied magnetic field
through the energy scale $\omega_B$. If one naively extends the behavior
of Eq.~(\ref{nu1dfinal-1}) to arbitrarily small frequencies, one would
have concluded that the TDOS goes to zero on the Fermi surface faster
than any power-law, if the renormalization of the conductivity
and the interaction strength at low energies is ignored.
This is in fact incorrect, because Eq.~(\ref{nu1dfinal-1}) is only
valid at intermediate frequencies. Since
Debye-Waller factor $e^{-W(it)}$ converges fast in the long time limit,
the low energy behavior of $\nu_{\rm q1D}(\omega)$ is actually described by
an asymptotic series expansion in powers of $\omega$. We find,
\be
\nu_{\rm q1D}(\omega)\approx \nu_0 \frac{|\omega|}{\omega_B} 
\sum_{n=0}^{\infty}(-
1)^{n}\frac{(4n+1)!}{(2n+1)!}\left(\frac{\omega}{\omega_B}\right)^{2n}.
\label{nu1dfinal-2}
\ee
The TDOS is therefore dominated by the linear term near zero bias,
\be
\nu_{\rm q1D}(\omega)\approx \nu_0 \frac{\omega}{\omega_B}=s\vov.
\label{nu1dfinal}
\ee
The magnetic field dependent slope is given by
\be
s=\frac{9\pi^4}{16 u^2}{1\over \rho_{xx}}(1+\lambda)^2,
\label{1dslop}
\ee
where $\rho_{xx}=\sxx(1+\gamma_H^2)$ is the dissipative resistivity.
Note that this result is valid at small frequencies such that
$\vov/\omega_B> e^{\omega_B/4\vov}$, i.e. for $\vov < 0.12\omega_B$.
In Fig.~3, we plot the TDOS obtained
by numerical integration of Eq.~(\ref{nu1d}) as a function of $\omega$.
The asymptotic linear pseudogap behavior of the TDOS at low bias is
shown in the inset. Therefore, we conclude that the
TDOS of a quasi-1D quantum Hall strip with reflecting edges exhibits
a linearly vanishing pseudogap near the Fermi level, with a slope
proportional to $\rho_{xx}^{-1}$ in the perturbative regime.
Within the SCBA, the values of both $\sxx$ and $\sxy$ 
at the center of the Landau levels 
are proportional to the Landau level index $N$.\cite{ando}
This leads to a Hall ratio $\gamma_H$ of order one and a longitudinal
resistivity $\rho_{xx}\sim 1/N\sim B$. We see that for a fixed 
interaction strength $u$, the slope is inversely proportional
to the magnetic field $s\sim 1/B$. It is interesting to remark
that bulk tunneling measurements under the quantum Hall conditions
using time-domain capacitance spectroscopy \cite{tunnelingexpt}
indeed reveals a linearly vanishing pseudogap TDOS with a slope
that scales with $1/B$. In addition, the measure slope of the tunneling
pseudogap oscillates weakly as a function of filling fraction,
which mimics the oscillatory behavior of the SCBA conductance.
However, although the experimental setup\cite{tunnelingexpt} 
allows screening of the Coulomb interaction by the metallic gates
(electrodes) such that the interactions may be short-range, it is not 
clear at the present if the the sample used can be 
qualified effectively as being quasi-one-dimensional. 
It is also interesting to note that the results for the quasi-1D
bulk TDOS in Eqs.~(\ref{nu1dfinal}) and (\ref{nu1dfinal-1})
depends strongly on the interaction strength $u$. This is in contrast 
to the case of tunneling into a single fractional quantum Hall edge
in the composite fermion description, where the TDOS was found to be
a power-law with an exponent that depends only weakly on the interaction 
strength.\cite{shytov}
\begin{figure}    
\center    
\centerline{\epsfysize=2.8in    
\epsfbox{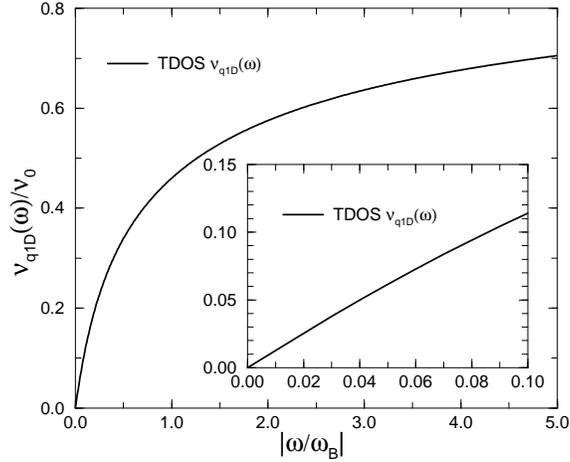}}
\begin{minipage}[t]{8.1cm}     
\caption{The TDOS in quasi-1D strips with edges obtained
from Eq.~(\ref{nu1d}) for short-range interactions.
The inset shows the asymptotic low energy behavior of
a linear pseudogap given by Eq.~(\ref{nu1dfinal}).
}   
\label{fig3}      
\end{minipage}     
\end{figure} 

\section{Discussions: Quantum Coulomb Gap and Dynamical Scaling of
Transition Width}

Understanding how interactions introduce new physics in the single-particle 
sector near the IQHT
is an important step towards a more complete understanding of
the interplay between disorder and correlation and its effects
on the transport properties in the quantum Hall regime.
A great part of this article is devoted to demonstrating
how various types of interaction-induced anomalies in the TDOS 
at low energy are likely to occur so long as the 2D conductivity is finite
which happens near the quantum Hall transitions. 
Our basic finding is that, in the presence of disorder, 
the range of validity for static screening of the Coulomb interaction
is very small, whereas at finite frequencies the diffusive dynamics
is to slow to effectively screen out the Coulomb interaction
at long distances. This leads to a Debye-Waller phase-delay factor
in the averaged single-particle Green's function that vanishes in the long-time
limit. As a result, the TDOS in the extended regime comes to 
resemble that in the localized regime, exhibiting a linearly vanishing
quantum Coulomb gap behavior.

It is important to emphasize the coexistence of the vanishing
Coulomb gap in the {\it tunneling} DOS with a finite thermodynamic DOS.
The double-log singularity
in Eq.~(\ref{nuomegacoul}), 
arising from the correlations of the single particle phases,
will not show up in the two-particle density-density correlation function
that determines the compressibility or the thermodynamic DOS
in the static limit. This point has recently been demonstrated explicitly in
the self-consistent Hartree-Fock theory,\cite{yang3} where it is shown
that the charge redistribution induced by a test charge inserted into 
the quantum Hall critical state is consistent with the presence of
a finite screening length. The finite
critical conductivity then implies that the uniform diffusion constant
must be finite. 

It is also important to understand how the depletion of TDOS
relates to the larger issue of dynamical scaling near the IQHT.
This is more challenging due to the possible existence of several
different time scales that control different dynamical processes:
charging, charge spreading, and inelastic phase breaking, etc.
While a linearly vanishing single-particle DOS in 2D does imply,
through the quasiparticle level spacing $\Delta\sim 1/L$,
{\it a} dynamical scaling exponent $z=1$, it has not been
shown how this $z$, which obviously controls the charging dynamics, 
is also the one that enters the conductivity scaling at finite temperatures
or frequencies in the transport measurements.

The conventional wisdom, at least for d.c. transport, has been that
the hopping transport in insulators
is determined by the single-particle DOS, whereas the diffusive transport
in metals is directly related to the thermodynamic DOS (compressibility).
Generalized to finite frequencies, this would imply that the dynamical 
aspects of the transport in insulators are controlled by that of charging, and
in metals by that of diffusive charge spreading. 
In ordinary disordered metals, this problem can be better quantified.
As noted by Finkel'stein \cite{finkelshtein} in the
calculation of the conductivity, there is a great degree
of cancelation between the corrections to the Green's functions (which
causes the anomalous behavior in the TDOS) and to the vertices.
Therefore the interaction effects that cause the depletion of the
the single-particle DOS $\nu(\omega)$ do not appear to influence 
directly the d.c. transport properties. 
At finite frequencies, the coupled scaling equations in
the RG calculation based on the nonlinear-$\sigma$ model involve
the conductivity, the interaction strength, and the frequency renormalization
$Z_\omega$ that enters in the diffusion kernel as
$1/(Dq^2-iZ_\omega\omega)$. 
Note that $Z_\omega=1$ corresponds to the non-interacting case and a
dynamical scaling exponent $z_\omega={\rm dim}[D]+2=d$ in d-dimensions.
In conventional Anderson-Mott
metal-insulator transitions in $d=2+\epsilon$ dimensions,\cite{bkrmp} 
the critical conductivity is zero.
The lack of quantum diffusion at the transition is accompanied
by the interaction induced frequency renormalization, 
i.e. $Z_\omega\sim L^{-\zeta}$.
For both the spin-scattering as well as the spin-polarized case,
one-loop calculations give $\zeta=\epsilon/2$. As a result, the dynamical
scaling exponent relevant for metallic transport departs from the 
noninteracting value, $z_\omega=d-\zeta$. 
It is very difficult to extend the same quantitative
analysis to the quantum Hall problem because the perturbative approach is 
no longer valid due to the presence of the topological $\theta$-term in 
the nonlinear-$\sigma$ model action. However, the fact that both $dn/d\mu$
and the conductivity are finite at the quantum Hall transition in 2D
ensures that the frequency in the diffusion propagator remains
unrenormalized and its associated exponent $z_\omega=2$.

The above analysis conveys a simple but important point, i.e. since
it is $Z_\omega$ and not the single-particle DOS that enters scaling 
and controls the dynamics of the diffusive transport 
from the metallic side, within the existing framework
\cite{finkelshtein,claudio}, it is natural to suspect that 
the suppression of the TDOS plays no role in the dynamical scaling
behavior of the conductivity. Therefore,
the linear Coulomb gap may appear not by itself an 
explanation as to why $z=1$ at the quantum Hall transitions.
In the following, we argue that it is indeed the interplay
between the quasiparticle inelastic dephasing (level broadening)
and the level spacing that controls the transition width,
contrary to common perceptions.

Since the relevant phenomenon here is transport, this state of affairs
naturally translates into the question of which exponent determines
the dephasing length $L_\varphi$. In the standard procedure, \cite{leerama}
the system is divided into $L_\varphi\times L_\varphi$ phase coherent
blocks. Transport within each block can be described by phase
coherent transport from the underlying noninteracting
theory and the relevant conductivity is given by the disorder-average
over the phase-coherent blocks. The outcome is that the scaling variable
for the conductivity becomes $L_\varphi/\xi$ in the presence of
interactions instead of $L/\xi$ in the noninteracting
case, where $L$ is the sample size and $\xi$ is the localization length,
\begin{equation}
\sigma_{xx}={e^2\over h}F\left({L_\varphi\over\xi}\right).
\label{scalingfunction}
\end{equation}
The scaling function $F(x)$ has the limiting behavior,
\be
F(x)=\left\{ \begin{array}{ll}
\sxxc, & x \to0, \\
0, & x \to\infty . \end{array} \right.
\label{scalingF}
\ee
The conducting critical regime at $L_\varphi \ll \xi$ and the insulating
regime at $L_\varphi \gg \xi$ are separated by a crossover 
at $L_\varphi\sim \xi$,
where the scaling variable in Eq.~(\ref{scalingF}) is of order one,
giving rise to a transition width
$\delta^*\sim L_\varphi^{-1/\nu_{\rm loc}}$.
Physically, the transition width can be viewed as the width
of the energy window of states whose localization length
exceeds the phase coherence length. In the language of
quantum critical phenomenon, the latter corresponds
to the width of the quantum critical region.
This is a generic property associated with 
the quantum critical point. The only peculiarity is that
on either side of the quantum Hall critical point the ground state
are insulators which, drawing analogy to quantum spin 
systems,\cite{sudip} are quantum disordered (see Fig.~4). 
The renormalized classical regimes (metallic phases) are absent. 

Now, we examine the conventional view of finding $L_\varphi$.
For a generic quantum phase transition \cite{sondhirmp,subir,wangetal}
the critical regime is characterized by the only time scale $\hbar/T$, and
thus the dephasing time $\tau_\varphi\sim 1/T$.
If quantum diffusion is all that matters, 
the associated length scale, i.e. the thermal diffusion length 
would be set by $L_T=(D\tau_\phi)^{1/2}\sim 1/T^{1/2}$
leading to a thermal exponent $z_T=2$.
Similarly, the length associated with a finite frequency is
$L_\omega=(D\hbar\omega)^{-1/2}$ such that the dynamic exponent 
$z_\omega=2$. The conventional approach is to identify $L_{T,\omega}$ 
with the quasiparticle
dephasing length $L_\varphi$ in the scaling function in 
Eq.~(\ref{scalingfunction}). As a result, the transition width in
this picture is determined by the crossover of length scales
set by $L_{T,\omega}\sim\xi$, indicated in Fig.~4 by the dashed
line, which leads to 
\be
\delta_{\rm would-be}^*\sim \left(T^{1/ z_T\nu_{\rm loc}},\omega^{1/ {z_\omega
\nu_{\rm loc}}}\right).
\label{deltastar}
\ee
As emphasized in the introduction, with the values $\nu\simeq2.3$ and
$z_T=z_\omega=2$, Eq.~(\ref{deltastar}) does not agree with the scaling
behavior of the transition width measured by transport 
experiments.\cite{hpwei,engel} From the theoretical point
of view, $\delta_{\rm would-be}^*$ would be the width of the phase 
coherent, diffusive metallic transport regime,
provided that Coulomb interaction effects are not too strong
to induce single-particle localization of the quasiparticle states.
\begin{figure}    
\center    
\centerline{\epsfysize=3.0in    
\epsfbox{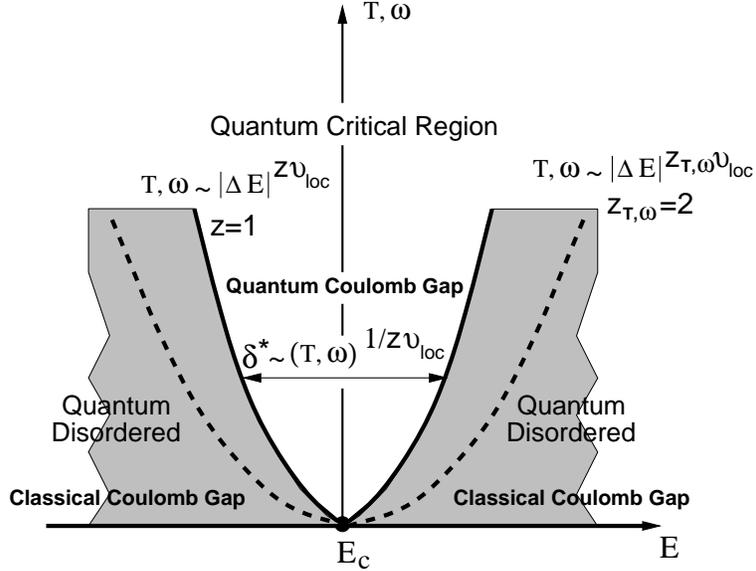}}
\begin{minipage}[t]{8.1cm}     
\caption{
Schematic phase diagram near the quantum critical point ($E_c$) of the IQHT.
The dashed line describes the ``would-be'' crossover between 
phase coherence, diffusive metallic transport of the
quantum critical region and the localized quantum disordered
regime at $L_{T,\omega}\sim\xi$. The associated thermal and
frequency exponents are $z_T=z_\omega=2$. The
solid line describes the true crossover from diffusive, metallic
to single-particle, insulator-like transport (shaded area) that takes
place when $L_\varphi\sim\xi$. The observed 
transition width, $\delta^*$, is narrower than $\delta_{\rm would-be}^*$
and has a scaling behavior controlled by the dynamical exponent
$z=1$, arising from the quantum Coulomb gap behavior.
}   
\label{fig4}      
\end{minipage}     
\end{figure}    
We now point out the problem with this picture which is commonly used
to describe metal-insulator transitions. The
use of $L_{T,\omega}$ as $L_\varphi$ in Eq.~(\ref{scalingfunction})
completely ignores
the important Mott physics in the single-particle sector, 
namely the tendency toward the single-particle insulator behavior
induced by Coulomb interaction. Physically, the inelastic dephasing time
is related to the interaction induced quasiparticle decay rate or level 
broadening $\Gamma\sim \hbar/\tau_\varphi$. The dephasing length $L_\varphi$,
on the other hand, can be determined only if the underlying transport 
mechanism is specified, ballistic or diffusive, insulating
or metallic. Clearly, diffusive metallic transport
is possible only if there is a significant overlapping of the quasiparticle
levels within $\Gamma$, i.e. the level broadening must be
larger than the interaction induced level spacing, $\Gamma > \Delta$.
In the opposite limit, $\Gamma < \Delta$,  the single-particle states 
are essentially gaped as a result of electron-electron interaction 
and the transport would be controlled by the localization in the 
single-particle sector similar to that in a Mott-insulator.
Diffusion would be impossible in this case and one expects variable 
range hopping to be the dominant mechanism of transport. 
It is therefore more appropriate to define the quasiparticle
dephasing length by the condition $\Gamma\sim \Delta(L_\varphi)$.
The presence of the quantum Coulomb gap, Eq.~(\ref{Coulomb2}),
in the critical regime implies that near the Fermi level, 
\be
\Delta(L_\varphi)\simeq {e^2\over\sqrt{C_Q}L_\varphi},
\label{deltac}
\ee
which leads to $L_\varphi\sim\tau_\varphi\sim1/T$
and the dynamical exponent $z=1$. Note that $L_\varphi/L_T\sim
T^{-1/2}$, i.e. $L_\varphi \gg L_T$ at low temperatures. However,
it is $L_\varphi$ that controls the crossover to the quantum
disordered insulating regime since when $L_\varphi > \xi$, the
level spacing within a $\xi\times\xi$ volume
becomes larger than the level broadening and the diffusive
metallic transport becomes impossible. It must be replaced by
hopping dominated transport similar to the Coulomb blockade regime
in quantum dots.
This part of the physics has been emphasized by Polyakov and Shklovskii
\cite{boris} and by Polyakov and Samokhin\cite{polyakov} in terms
of the classical Coulomb gap.

Using this $L_\varphi$, the scaling 
function in Eq.~(\ref{scalingfunction}) now describes 
the true crossover line, set by $L_\varphi\sim\xi$ (the solid line
in Fig.~4), that separates the diffusive, metallic transport
from the single-particle, insulator-like transport.
The scaling behavior of the width of the critical conducting regime
is therefore given by
\be
\delta^*\sim (T,\omega)^{1/z\nu_{\rm loc}},
\ee
with $z=1$ as observed in transport measurements. 
In this theory, the presence
of the quantum Coulomb gap behavior is central to the emergence
of the $z=1$ scaling of the transition width. It eliminates
the difficulty associated with
invoking the classical Coulomb gap or the bare charging energy due to the
unscreened Coulomb interaction \cite{boris,polyakov} which is only
valid deep in the insulating regime.

As shown in Fig.~4, the single-particle
DOS obeys the quantum Coulomb gap behavior in the quantum critical
regime, whereas deep in the quantum disordered, insulating regime, 
it is expected that the Coulomb interaction reinstates the classical 
Coulomb gap of Efros and Shklovskii.\cite{ccgap} 
Thus it is remarkable that in the presence of
Coulomb interactions, the crossover between the quantum critical
and the quantum disordered regimes is accompanied by a crossover in
the behavior of the TDOS --- from the quantum to the classical Coulomb gap.
Comparing Eqs.~(\ref{classical}) and (\ref{Coulomb}), we see that the
crossover is simply described by a crossover in the slope of the 
linear gap near the Fermi level. Such a crossover should, in principle,
be detectable experimentally by sitting at a fixed distance to
the critical point of the transition. In this case, as the temperature
or frequency is lowered, one should observe a linearly
vanishing gap with an initial nonuniversal slope that turns into a universal
number in the low temperature/frequency limit.

We emphasize that the linear {\it quantum} Coulomb gap behavior results from
the combined effects of (i) two-dimensionality, 
(ii) long-range Coulomb potential, and (iii) quantum diffusion, {\it i.e.} 
a finite conductivity at $T=0$. It is expected to pertain
to other metal-insulator transitions in 2D amorphous electron systems,
provided that the critical conductivity is finite.
The physics discussed here is quite generic of the
2D disordered metal-insulator quantum critical point.
A recent example is the 2D $B=0$ metal-insulator 
transition.\cite{krav} In that case, Fig.~4 needs to be modified
to include the renormalized classical, i.e. the metallic region.
It is our hope that the present work will stimulate further
experimental investigations on the nature of dynamical scaling
in the quantum Hall effect and in other metal-insulator transitions.

\section{Acknowledgments} 

The authors thank X. Dai, M. Fogler, S. Girvin, A. Houghton, B. Huckstein, 
A. Kamenev, H.-Y. Kee, Y.~B. Kim, D.-H. Lee, A. MacDonald, 
B. Shklovskii, and E. Yang for useful discussions.
Z.W. thanks Aspen Center for Physics and the Institute for Theoretical
Physics at UCSB for hospitality. This work is supported in part 
by DOE Grant No. DE-FG02-99ER45747 and an award from
Research Corporation.

\appendix

\section{The Matsubara sum}

In this appendix, we carry out the discrete frequency sum in the phase delay
given in Eq.~({\ref{finaldw}):
\be
W(\tau)=\frac{1}{\beta}\sum_{n}[1-e^{i\omega_n\tau}] f(|\omega_n|),
\ee
where $\omega_n=2\pi n /\beta$, $n=0,\pm1,\pm2\dots$ 
is the boson Matsubara frequency and
\be
f(|\omega_n|)= \left[\frac{1}{|\omega_n|+Dq^2}\right]^2 
\frac{v(q)}{1+ v(q) \nu_0 \frac{Dq^2}{|\omega_n|+Dq^2}}.
\ee
To perform the Matsubara sum, we first separate the positive and 
the negative frequencies by writing
\be
W(\tau)=W^{+}(\tau)+W^{-}(\tau),
\ee
where
\bea
W^{+}(\tau)&=&\sum_{n>0}\frac{1}{\beta}[1-e^{i\omega_n\tau}]f(\omega_n),
\\
W^{-}(\tau)&=&\sum_{n<0}\frac{1}{\beta}[1-e^{i\omega_n\tau}]f(-\omega_n).
\eea
Next, we define a function on the complex plane,
\be
F^{\pm}(z)=\frac{1-e^{z \tau}}{e^{\beta z}-1}f(\mp iz).
\label{fz}
\ee
We consider the integrals of $F(z)$ along contours in the upper 
($C_1+C_2$) and the lower ($C_3+C_4$) half-plane as
shown in Fig.~5. The results are given by, respectively:
\bea
\oint_{C_1+C_2}F^{+}(z)dz& =&\frac{2\pi 
i}{\beta}\sum_{n>0}[1-e^{i\omega_n\tau}] f(\omega_n)+2\pi 
i \times \mbox{residues from}\; f(-iz)\non\\
\oint_{C_3+C_4}F^{+}(z)dz& =&\frac{2\pi 
i}{\beta}\sum_{n<0}[1-e^{i\omega_n\tau}]f(-\omega_n)+2\pi 
i \times \mbox{residues from}\; f(iz)
\eea
It can be shown straightforwardly that the integrals of $F(z)$ 
along both semi-circles  $C_2$ and $C_4$ 
(with $|z|=R$) vanish at infinite radius $R\to \infty$, provided that
$0<\tau<\beta$. Since residues of $f(-iz)$ lie
in the lower half-plane while those of $f(iz)$
lie in the upper half plane, they do not contribute to the contour 
integrals as we defined. Therefore, summing up the integrals along
$C_1$ and $C_3$, we obtain,
\be
W(\tau)=\int_{-\infty}^{+\infty}d\epsilon \frac{1-
e^{\tau\epsilon}}{e^{\beta\epsilon}-1}\frac{1}{2\pi i}[f(-i\e)-f(i\e)].
\label{wsumed}
\ee
Note that since the fermionic Green's functions are anti-periodic in
$\tau$, i.e. $G(\tau+\beta)=-G(\tau)$, it implies that, through
Eq.~(\ref{finalG}), a periodic phase factor $W(\beta+\tau)=W(\tau)$ which
is indeed satisfied by Eq.~(\ref{wsumed}).
\begin{figure}    
\center    
\centerline{\epsfysize=2.8in    
\epsfbox{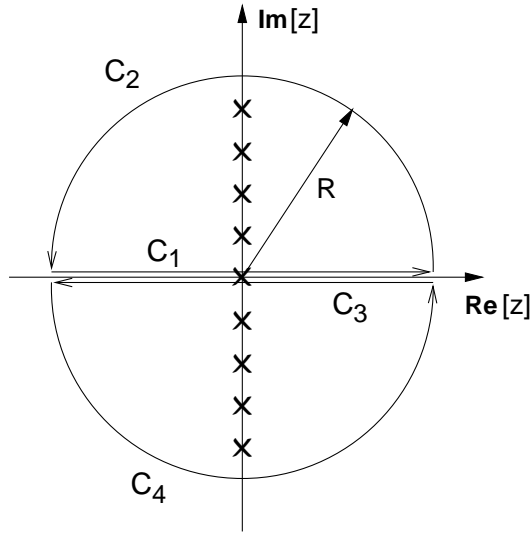}}
\begin{minipage}[t]{8.1cm}     
\caption{
The contours of integration for the functions
$F^{\pm}(z)$ defined in Eq.~(\ref{fz}). The crosses indicate the
locations of the poles at Matsubara frequencies
$\omega_n=2\pi n/\beta$.
}   
\label{fig5}      
\end{minipage}     
\end{figure}    
\section{Analytical continuation}

In this appendix, we describe one of the technical subtleties encountered
when taking the analytical continuation in Eq.~(\ref{analytic}).
We show how to obtain $G(\omega+i0^+)$ directly from the time ordered 
imaginary-time Green's function $G(\tau)$ by analytical continuing
$\tau\to it+0^+$. We begin with the Fourier transform of the 
fermion Green's function,
\be
G(i\omega_n)=\int_0^{\beta} d\tau 
e^{i\omega_n\tau}G(\tau),
\label{mastu}
\ee
which satisfies the anti-periodic
boundary condition $G(\tau+\beta)=-G(\tau)$. As a result
$G(i\omega_n)$ is non-zero only for odd Matsubara frequencies, 
i.e. for $\omega_n=\frac{(2n+1)\pi}{\beta}$.

To perform the integration in Eq.~(\ref{mastu}) and the analytical
continuation to real frequency, we extend $\tau$ to the complex
$z$-plane with ${\rm Re}[z]=\tau$ and $i{\rm Im}[z]=it$.
We seek to analytically continue the integral in the segment bounded
by $(0,\beta)$ on the real axis to integrals along the vertical axis
at $\tau=0,\beta$. To this end,
consider the closed-path integral along the contour shown in
Fig.~6, chosen to lie in the upper half-place for $\omega_n>0$. 
Since $G(\tau)$ is non-analytic at $\tau=0,\beta$, the vertical
segments of the contours are shifted infinitesimally such that
$0<{\rm Re}z <\beta$.
The analytical continuation is possible when $G(z)$ is analytic
and has no poles encircled by the contour,
\be
\oint dz e^{i\omega_n z} G(z)=0,
\label{contour}
\ee
Since the integral along the $\vert z\vert\to\infty$ segment
of the contour in Fig.6 vanishes for $\omega_n>0$, we have
\be
G(i\omega_n)=i\int_0^{\infty} d t e^{-\omega_n t}[G(it+0^+)-
G(it+\beta)].
\ee
Using the anti-periodic property, we obtain
\be
G(i\omega_n)=2i\int_0^{\infty} d t e^{-\vert\omega_n\vert t}G(it+0^+),
\ee
where we have included the result for $\omega_n<0$, in which case,
the integration contour was chosen to lie in the low half-plane.

Next we take the analytical continuation in frequency:
$i\omega_n\to\omega+i0^+$ and obtain,
\be
G(\omega+i\dl)=2i\int_0^{\infty} d t e^{i\omega t}G(it).
\simeq 2i \int_{\tau_0}^\infty e^{i\omega t}\ov{G}_0 e^{-W(it)},
\label{gomega}
\ee
where we have used Eq.~(\ref{finalG}) for the Green's function $G(it)$
in our semi-classical phase approximation. Substituting the
expression of the SCBA Green's function $\ov{G}_0$ in Eq.~(\ref{scbag}),
we obtain the TDOS at finite temperatures,
\be
\nu(\omega)=-\frac{1}{\pi} {\rm Im} G(\omega+i\dl)= \nu_0 
\frac{2}{\beta}\int_{\tau_0}^{\infty}{\sin(|\omega| t)\over
\sinh(\pi t/\beta)}e^{-W(it)}.
\ee
Note that an overall factor stemming from the Fermi distribution
function has not been included in the 
definition of $\nu(\omega)$ at finite temperatures, since it is, at any
rate, unimportant at low temperatures.
\begin{figure}    
\center    
\centerline{\epsfysize=2.8in    
\epsfbox{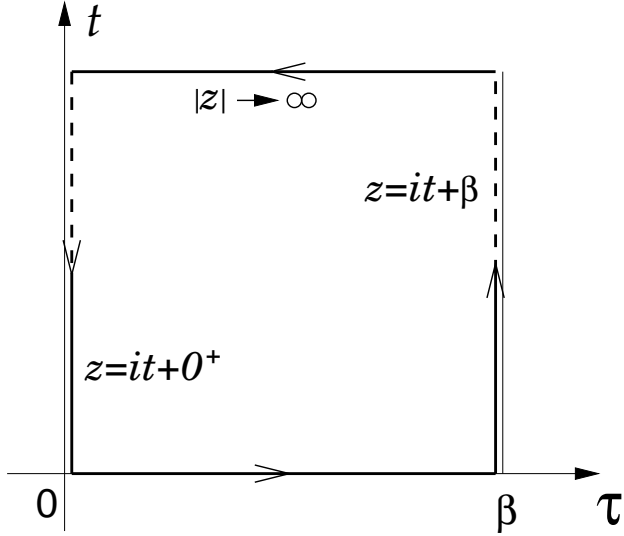}}
\begin{minipage}[t]{8.1cm}     
\caption{
The contour of integration used in Eq.~(\ref{contour})
for carrying out the analytical continuation. 
The real axis corresponds to the imaginary time $\tau$, 
whereas the imaginary axis is labeled by the real time $t$.
}   
\label{fig6}      
\end{minipage}     
\end{figure}    

\section{Six integration regions} 

In this appendix, we perform the integrations over the
frequency and wave-vector in Eq.~(\ref{dwintegral}) to obtain the
phase-delay factor $W(it)$.
We consider the interaction potential of the 
general form $ v(q)=u\frac{1}{q^{2-p}}$ in momentum space 
or $v(r)=u/r^p $ in real space. We can rewrite Eq.~(\ref{dwintegral} as
\be
W(it)=\int_{1/t}^{\infty}d \omega \frac{ d \omega }
{2\pi i} \int\frac{d q^2}{4\pi^2} Q(q,\omega), 
\ee
where the integrand with $\kappa^{2-p}=\nu_0 u$ is given by
\be
Q(q,\omega)=\frac{2u\omega (2Dq^2 +D q^p 
\kappa^{2p})}{[(Dq^2)^2+\omega^2][(Dq^2+Dq^p \kappa^{2-p})^2+
\omega^2] q^{2-p}}.
\label{Qintegrand}
\ee
There are six characteristic integration regions differed by the
ordering of $Dq^2$, $Dq^p\k^{2-p}$ and $\omega$. The details of the
integration are discussed below. We find  for all $p<2$ 
the leading contribution is the double-log term appeared in
the long-range Coulomb case (p=1) in Section IV.A.
For $p>2$ the leading contribution is of the form $W_{\infty}\sim 1/t^{\dl}$ 
with $\dl>0$, and $\dl\rightarrow 0$ as $p\rightarrow 2$. 

\subsection{The regime $Dq^2<\omega<D q^p \kappa^{2-p}$  }

The condition $Dq^2<\omega<D q^p \kappa^{2-p}$ requires
$\omega< D\k^2$ for $p<2$; and
$\omega>D\k^2$ for $p>2$. In this case, the integrand $Q$ in 
Eq.~(\ref{Qintegrand}) can be approximated by,
\be
Q(p,\omega)\approx \frac{2}{D\nu_0 \omega q^2}.
\label{Q1}
\ee
We get for $p<2$,
\be
\int {d^2 q \over4\pi^2}
Q(q,\omega)=-\frac{1}{2\pi}\frac{1}{D\nu_0}\frac{2-p}{p}\log 
{\omega\tau_s}
\ee
where we have introduced $\tau_s$ via $1/\tau_s=D\k^2$. Performing
the remaining $\omega$ integral, we obtain
\be
\int_{1/t}^{1/\tau_0}\frac{ d\omega }{2\pi} 
\int \frac{ d^2 q
}{4\pi^2} Q(q,\omega)=\frac{1}{8\pi^2}\frac{1}{D\nu_0}
\frac{2-p}{p}\log{t/\tau_0}\log{t/\tau_1},
\ee
where $\tau_1^2=\tau_0/D\k^2$.

For $p>2$, due to the requirement that  $\omega>D\k^2$, this regime does 
not have any time-dependent contribution in the long-time limit when 
$t>\tau_s$.

\subsection{The regime $Dq^2<D q^p \k^{2-p}<\omega$}

This regime requires $q< q_0(\omega)
\equiv{\rm min}[\k, (\omega\tau_s)^{1/p}\k]$
for $p<2$ and $\k<q<(\omega\tau_s)^{1/p}\k$ for $p>2$.
The latter case requires $\omega>1/\tau_s$. This means that 
for $p>2$, this regime does not contribute in the
long-time limit in a time-dependent way. 

For $p<2$, we have
\be
Q(q,\omega)\approx\frac{2}{\omega^3}\nu_0 u^2 D q^{2p-2},
\label{Q2}
\ee
such that
\be
\int {d^2 q\over4\pi^2} Q(q,\omega)=\frac{1}{\pi\omega^3}\nu_0u^2 
D\int_{0}^{q_0(\omega)}dq q^{2p-
1}=\frac{1}{\pi}\nu_0u^2D 
\frac{q_0^{2p}(\omega)}{2p}.
\ee
Usually $t>\tau_0>\tau_s$, therefore $\omega<\tau_s^{-1}$. We get
 \be
\int_{1/t}^{1/\tau_0}\frac{ d\omega }{2\pi} 
\int \frac{ d^2 q 
}{4\pi^2} Q(q,\omega)= 
\frac{1}{2\pi^2}\nu_0u^2D\frac{1}{2p} \k^{2p}\tau_s^2\log 
(t/\tau_0)=\frac{u}{2\pi^2}\frac{1}{2p}\tau_s \k^p \log(t/\tau_0).
\ee 
This single-log term is subleading when compared to the leading
double-log contribution in regime 1 in the long-time limit.

\subsection{The regime $Dq^pk^{2-p}<Dq^2<\omega$}

In this case, we have
\be
Q(q,\omega)\approx \frac{4 Du}{ \omega^3}q^p.
\label{Q3}
\ee
The limits for the $q,\omega$ integrals are
$\k<q<\sqrt{\omega/D}$, $\omega>D\k^2=1/\tau_s$ for $p<2$; and
$q<\min[\k,\sqrt{\omega/D}]$ for $p>2$. 

The $p<2$ case is of no interest in this regime
since the lower cut-off of the 
frequency integral is $1/\tau_s$ and is time-independent in the limit 
$t>\tau_s$.    

We now discuss the $p>2$ case. For most of the physical systems the 
mean-free path $\ell$ is greater than the screening length $\k$, 
therefore $\min\{\k, \sqrt{\omega/D}\}= \sqrt{\omega/D}$. The
integrals can be carried out according to,
\bea
\int {d^2 q \over4\pi^4}
Q(q,\omega)&=&\frac{2Du}{\pi}\frac{1}{\omega^3}\int_0^{\sqrt{\omega
/D}}qdq q^p=\frac{2u}{\pi}\frac{1}{p+2}\frac{1}{D^{p/2}}(\omega)^{p/2-
2},
\\
\int_{1/t}^{1/\tau_0}\frac{ d\omega }{2\pi} 
\int \frac{ d^2 q 
}{4\pi^2}Q(q,\omega)&=&\frac{u}{\pi^2D^{p/2}}\frac{1}{p+2}\frac{2}{p-
2}\left[\frac{1}{\tau_0^{p/2-1}}-\frac{1}{t^{p/2-1}}\right].
\eea

\subsection{The regime $Dq^p\k^{2-p}<\omega<Dq^2$}

In this regime, we have
$$
Q(q,\omega)\approx\frac{4\omega u}{D^3}\frac{1}{q^{8-p}}
$$

The limits of the integrations are:
$(\omega/D)^{1/2}<q<\k (\omega\tau_s)^{1/p}$; 
$\omega>1/\tau_s$ for $p<2$ and $\omega<1/\tau_s$ for $p>2$.  
The $p<2$ case only produces
time-independent contributions in the limit $t>\tau_s$.

For the case of $p>2$, it is straightforward to obtain
\bea
\int {d^2 q\over4\pi^2} Q(q,\omega)&=&\frac{2u 
\omega}{D^3\pi}\int_{(\omega/D)^{1/2}}^{\k(\omega\tau_s)^{1/p}}dq 
\frac{1}{q^{7-p}}=\frac{2u}{D^3\pi}\frac{1}{6-
p}\left(\frac{1}{\omega^{2-p/2}D^{p/2-3}}-
\frac{1}{\omega^{6/p-2}\tau_s^{6/p-1}\k^{6-p}}\right),
\\
\int_{1/t}^{1/\tau_0}\frac{ d\omega }{2\pi} 
\int \frac{d^2 q}{4\pi^2}  
Q(q,\omega)&=&\frac{u}{\pi^2}\frac{1}{6-p}\left\{\frac{2\tau_0}{(p-
2)\ell^p}[1-
(\tau_0/t)^{p/2-1}]-\frac{p\tau_s \k^p}{(3p-6)(\k \ell)^6 
(\tau_s/\tau_0)^{6/p}}[1-
(\tau_0/t)^{3-6/p}]\right\},
\eea
where $\ell=(2D\tau_0)^{1/2}$ is the mean free path.

\subsection{The regime $\omega<Dq^p\k^{2-p}<Dq^2$}

In this regime, the integrand $Q$ in Eq.~(\ref{Qintegrand}) can be
approximated by
\be
Q(q,\omega)\approx \frac{4u\omega}{D^3}\frac{1}{q^{8-p}}.
\label{Q5}
\ee
For $\omega<1/\tau_0<1/\tau_s$, the limits of the integrations are
$q>\k$ for $p<2$ and $(\omega\tau_s)^{1/p}\k<q<\k$ for $p>2$. 

In the case of  $p<2$, we have
\bea
\int {d^2 q\over4\pi^2} Q(q,\omega)=\frac{2u 
\omega}{D^3\pi}\int_{\k}^{\infty}dq 
\frac{1}{q^{7-p}}&=&\frac{2u\omega}{D^3\pi}\frac{1}{6-p}
\frac{1}{\k^{6-p}}, \\
\int_{1/t}^{1/\tau_0}{d\omega\over2\pi}\int 
{d^2q\over4\pi^2}
Q(q,\omega)&=&\frac{u\tau_s^3 \k^p}{2\pi^2 \tau_0^2(6-p)}[1-
(\tau_0/t)^2].
\eea
On the other hand, for $p>2$, we have
\be
\int {d^2q\over4\pi^2} Q(q,\omega)=\frac{2u
\omega}{D^3\pi}\int_{\k}^{\infty}dq 
\frac{1}{q^{7-p}}=\frac{2u\omega}{D^3\pi}\frac{1}{(6-p)\k^{6-
p}}\left[\frac{1}{(\omega\tau_s)^{(6-p)/p}}- 1\right],
\ee
leading to
\bea
\int_{1/t}^{1/\tau_0}{d\omega\over2\pi}\int 
{d^2 q\over4\pi^2} Q(q,\omega)=&&
\frac{u}{\pi^2}\frac{1}{6-p}\frac{p\tau_s \k^p}{(3p-6)(\k \ell)^6 
(\tau_s/\tau_0)^{6/p}}[1-
(\tau_0/t)^{3-6/p}] \non\\
&&-\frac{u\tau_s^2 \k^p}{2\pi^2 
\tau_0^2(6-p)}[1-(\tau_0/t)^2].
\eea

\subsection{The regime $\omega<Dq^2<Dq^p\k^{2-p}$}

Finally, in regime six, we have
\be
Q(q,\omega)\approx\frac{2\omega}{\nu_0 (Dq^2)^3}.
\label{Q6}
\ee
The limits of the integrations are
$(\omega/D)^{1/2}q<\k$ for $p<2$ and
$q>\k$ for $p>2$.

In the $p<2$ case, integrals give
\bea
\int {d^2q\over4\pi} Q(q,\omega)&=&\frac{1}{4\pi}\frac{\omega}{\nu_0 
D^3}\left[\frac{D^2}{\omega^2}-\frac{1}{\k^4}\right], \\
\int_{1/t}^{1/\tau_0}{d\omega\over2\pi}\int 
{d^2q\over4\pi^2} Q(q,\omega)&=&
\frac{1}{8\pi^2 \nu_0 D}\left\{\log(t/\tau_0)-
\frac{1}{2D^2\k^4}[1/\tau_0^2-
1/t^2]\right\}.
\eea

The  $p>2$ case gives, on the other hand,
\be
\int_{1/t}^{1/\tau_0}{d\omega\over2\pi}\int 
{d^2q \over4\pi^2}Q(q,\omega)=
\frac{1}{16\pi^2 \nu_0D}\tau_s^2[1/\tau_0^2-1/t^2].
\ee

Note that the $\delta$-potential considered in Section IV.B
corresponds to the $p=0$ case. The result in Eq.~(\ref{W-short})
can be obtained from either the $p>2$ case or the $p<2$ case
by taking the limit $p\rightarrow 2$  using 
$\lim_{x\rightarrow 0} \frac{1}{x}(1-y^{x})=-\ln y$ and 
$\lim_{p\rightarrow 2}(p-2)\ln (\tau_1)=\ln (\nu_0u) $.



\begin{thebibliography}{99}
\bibitem{notehf} Note that upon loop-corrections, the residual Coulomb
interaction beyond the HF theory can be either marginally relevant, 
in which case a change in $\nu$ is still expected but would be difficult
to observe, or marginally irrelevant
such that the HFFP becomes the stable fixed point of the IQHT in the
presence of Coulomb interactions.
\end{thebibliography}
\end{document}